\documentclass[twocolumn,twoside,trackchanges]{aastex701}

\usepackage{appendix}
\usepackage{float}
\usepackage{tabularx}
\usepackage{amsmath, amssymb}
\usepackage{hyperref}
\usepackage{booktabs}
\usepackage{longtable}

\begin{document}

\title{Chemodynamical Study and Evidence for Hot Compact Companions to Blue Straggler Stars in Berkeley 31}

\correspondingauthor{D. Bisht}
\email{devendrabisht297@gmail.com}  

\author[orcid=0000-0001-7940-3731, sname=Çınar, gname=Deniz Cennet]{Deniz Cennet Çınar}\thanks{E-mail: denizcennetcinar@gmail.com}
\affiliation{Programme of Astronomy and Space Sciences, Institute of Graduate Studies in Science, Istanbul University, Istanbul, 34116, Turkey}
\email{denizcennetcinar@gmail.com}

\author[orcid=0000-0002-8988-8434, sname=Bisht, gname=D.]{D. Bisht}
\affiliation{Indian Centre for Space Physics
466, Barakhola, Singabari Road, Netai Nagar, Kolkata, West Bengal, 700099}
\email{devendrabisht297@gmail.com}

\author[orcid=0000-0001-7359-3300, sname=Jiang, gname=D.]{Ing-Guey Jiang}\thanks{E-mail: jiang@phys.nthu.edu.tw}
\affiliation{Department of Physics and Institute of Astronomy, National Tsing-Hua University, Hsinchu 30013, Taiwan}
\email{jiang@phys.nthu.edu.tw}

\author[sname=Belwal, gname=K.]{K. Belwal}
\affiliation{Indian Centre for Space Physics 466, Barakhola, Singabari Road, Netai Nagar, Kolkata, West Bengal, 700099}
\email{kuldeepbelwal1997@gmail.com}

\author[0000-0003-3510-1509]{Sel\c{c}uk Bilir} \affiliation{Istanbul University, Faculty of Science, Department of Astronomy and Space Sciences, 34119, Beyaz\i t, Istanbul, Turkey} \email{sbilir@istanbul.edu.tr}

\begin{abstract}
We present a detailed study of the open cluster Berkeley~31 using Gaia~DR3 astrometry, Gaia-ESO Survey DR5.1 spectroscopy, and multi-wavelength photometry. We identify 329 probable cluster members from their proper motions and trigonometric parallaxes and use them to examine the cluster's fundamental parameters, stellar populations, and Galactic orbit. The cluster is located at a distance of $6024\pm181$~pc, with a color excess of $E(B-V)=0.148\pm0.047$~mag, and has a mean metallicity of [Fe/H] $=-0.36\pm0.04$~dex and $\langle[\alpha/{\rm Fe}]\rangle=0.21\pm0.04$~dex. Isochrone fitting gives an age of $3.7\pm0.2$~Gyr, consistent with the independent chemical-clock estimate of $3.72\pm1.42$~Gyr. The systemic radial velocity is $57.89\pm0.08$~km~s$^{-1}$. We find that about $38\%$ of the MS stars are unresolved binary candidates with $q\gtrsim0.5$. We identify 62 BSSs and 2 YSS candidates, including 48 BSSs and both YSSs not previously reported in the literature. We constructed multi-wavelength SEDs for these stars, including \textit{Swift}/UVOT data for 28 BSSs. Two BSSs show clear ultraviolet excesses and are consistent with compact hot companions, one with a hot WD and the other with a pre-ELM WD. The present orbital parameters give a guiding radius of $12.29\pm0.29$~kpc and an eccentricity of $\sim 0.11\pm0.01$. Comparison with the estimated chemical birth radius indicates a moderate outward shift of the guiding center, with the present orbital position also contributing to the radial displacement. These results consistently describe Berkeley~31 as an intermediate-age, metal-poor outer-disk cluster and place its stellar populations in the context of its Galactic dynamical history.
\end{abstract}

\section{Introduction}
\label{sec:intro}

Open clusters (OCs) are useful tracers of Galactic disk formation and evolution. Their stars are formed from the same molecular environment and are approximately coeval, making OCs valuable for studying stellar evolution and the chemical structure of the Milky Way. Their distribution across the Galactic disk provides measurements of metallicity and elemental abundances over a wide range of ages and Galactocentric distances \citep[e.g.,][]{Friel1995, Friel2002, Magrini2009, Carrera2011, Donor2020, Yontan2026}.

The Gaia mission has greatly improved the study of OCs by providing precise astrometry for large numbers of stars. Parallaxes and proper motions provide a homogeneous basis for identifying cluster members and separating them from the surrounding field \citep{GaiaCollaboration2016, GaiaCollaboration2018, GaiaCollaboration2021, GaiaCollaboration2023}. Reliable membership samples are essential for determining cluster distances, structural parameters, and color-magnitude diagrams (CMDs). When combined with radial velocities (RVs), Gaia astrometry also provides the phase-space information needed to study cluster kinematics and Galactic orbits \citep[e.g.,][]{Soubiran2018, Tarricq2021, Yontan2022, Yontan2023a, Yontan2023b, Bilir2026}.

Determining cluster parameters from CMDs is not straightforward because age, distance, extinction, and metallicity are partly degenerate. Different combinations of these parameters can produce similar CMD morphologies, making a unique solution difficult to identify \citep[e.g.,][]{vonHippel2006, Bilir2006a, Bilir2006b, Bilir2010, Bilir2016, Yontan2019}. In practice, this problem can be further complicated by field-star contamination, differential extinction, photometric uncertainties, and unresolved binaries, all of which can broaden or distort the observed cluster sequences \citep{Bostanci2015, Bostanci2018, Ak2016, Banks2020, Yontan2021, Akbulut2021}. Consequently, reliable determination of the fundamental parameters generally requires the CMD morphology to be considered together with the observational and astrophysical effects that shape it \citep{Karaali2011, Yontan2015,  Bossini2019, Gokmen2023,  Cakmak2024}. This is particularly important for age estimates, since changes in distance, extinction, or metallicity can reproduce some of the effects of age on the CMD. Bayesian isochrone fitting combined with Markov Chain Monte Carlo (MCMC) sampling provides a useful framework for exploring these correlated parameters and estimating their uncertainties simultaneously \citep{Tanik2025, Karagoz2025, Tasedemir2026, Elsanhoury2026}. However, MCMC does not remove the underlying degeneracies; the inferred age still depends on how well the other parameters are constrained. Independent estimates of distance, extinction, and chemical composition can therefore restrict the allowed parameter space and lead to more robust age determinations \citep[e.g.,][]{Cinar2026d, Cinar2026c}.

Spectroscopic data can provide these additional constraints. Surveys such as the Gaia-ESO Survey (GES) \citep{Gilmore2022, Randich2022} provide homogeneous measurements of elemental abundances, stellar atmospheric parameters, and RVs \citep{Hourihane2023}. Quantities such as metallicity, effective temperature, surface gravity, and RV can be constrained from stellar spectra for individual stars, rather than inferred solely from the CMD. Together with Gaia-based membership information, these measurements allow the cluster metallicity and systemic RV to be determined independently of the CMD. Fixing or tightly constraining the chemical composition from spectroscopy also prevents metallicity from compensating for changes in age during the isochrone fitting. Such constraints reduce the freedom in the isochrone fit and make the resulting cluster age less sensitive to the degeneracies among the fitted parameters.

The resulting cluster parameters also provide a basis for studying the evolution of its stellar populations. Blue straggler stars (BSSs), which lie above the main-sequence turn-off (MSTO) in a CMD, are particularly useful in this context because their properties can retain information about binary interactions and other processes that occurred during the cluster's evolution. BSSs are thought to form through channels that may include binary mass transfer and stellar mergers, with their relative importance depending on the properties of the host system \citep{Schneider2026}. In some clusters, hot compact companions can be identified from ultraviolet excess emission that cannot be explained by the cool BSS component alone. Ultraviolet observations have therefore been used to search for white dwarf (WD) and pre-WD companions to BSSs and to identify systems that may have undergone past binary mass transfer \citep[e.g.,][]{Knigge2009, Geller2012, gosnell2014detection}. Such companions provide an additional observational link between the present-day BSS population and its evolutionary history.

Accurate cluster parameters are also needed to place the cluster in its Galactic dynamical context. The present-day Galactocentric position of an OC does not necessarily represent its formation location, since stellar orbits can change over time through processes associated with Galactic dynamics \citep{Sellwood2002, Roskar2008, Minchev2011}. Distances, RVs, and proper motions can be used to derive Galactic space velocities and orbital parameters, while orbit integration provides estimates of quantities such as guiding radius, eccentricity, and maximum vertical height \citep{Bovy2015, Anders2017, Spina2021, Viscasillas2022}. These quantities help place the observed stellar populations in the cluster's broader dynamical history and can be used to investigate whether its present location is consistent with its orbital evolution.

Berkeley 31 (Biurakan 7,  [KPS2012] MWSC 1032, C~0654+083, and OCl 513; $\alpha_{2000} = 06^\mathrm{h}57^\mathrm{m}37^\mathrm{s}.4$, $\delta_{2000} = +08\degr17\arcmin06\arcsec$; $l=206\degr.2397$, $b=+5\degr.1332$) is an intermediate-age, metal-poor OC located in the outer Galactic disk toward the Galactic anti-center. Early photometric observations by \citet{Guetter1993} yielded an age of approximately 8 Gyr, a distance of $5.24\pm0.50$ kpc, a reddening of $E(B-V)=0.13$ mag, and a metallicity of [Fe/H]=-0.40 dex. Using low-resolution spectra of 24 stars, \citet{Friel2002} identified 17 probable cluster members and derived a mean metallicity of [Fe/H]=-0.40 dex (rms = 0.16 dex) and a mean radial velocity of +61 km s$^{-1}$ (rms = 15 km s$^{-1}$). The relatively large radial velocity dispersion, however, raised questions about the reliability of cluster membership. Using deeper photometry and synthetic color-magnitude diagrams based on different stellar evolutionary models, \citet{Cignoni2011} obtained a considerably younger age of 2.3-2.9 Gyr, with reddening values of $E(B-V)=0.095–0.185$ mag and distance estimates that depend on the adopted models and metallicity. These studies established Berkeley 31 as an important outer-disk cluster, but also revealed substantial differences in its inferred fundamental parameters, particularly its age and distance. Given its location and subsolar metallicity, Berkeley 31 provides an opportunity to investigate the chemical properties and orbital evolution of an evolved stellar population in the outer Galactic disk.

In this study, we present a comprehensive analysis of Berkeley 31 by combining Gaia DR3 astrometry, GES DR5.1 spectroscopy, and ultraviolet observations to examine its fundamental parameters, chemical composition, stellar populations, and Galactic dynamics. We first establish cluster membership using Gaia astrometry and combine the resulting membership information with spectroscopic constraints on metallicity and systemic radial velocity to determine the cluster parameters. We then investigate its blue straggler star (BSS) population using available \textit{Swift}/UVOT observations, searching for ultraviolet excesses that may indicate the presence of hot compact companions associated with past binary mass transfer. Finally, we use the derived kinematic properties to examine the Galactic orbit of Berkeley 31 and assess its possible radial migration history. By bringing these measurements together, we aim to provide a more consistent characterization of the cluster and place its stellar populations in the context of the chemical and dynamical evolution of the outer Galactic disk.

\section{Data}
\label{sec:data}

We first investigate the fundamental properties and stellar population of Berkeley~31 using Gaia DR3 photometric and astrometric data, complemented by spectroscopic information from the Gaia-ESO Survey. We then adopt a multiwavelength approach to characterize the BSS candidates, combining photometric data across the ultraviolet, optical, and infrared regimes with astrometric and spectroscopic information available for cluster members.

\subsection{Gaia DR3 Astrometry and Photometry}
\label{subsec:astrometry_data}

The astrometric and photometric data used in this study were obtained from the third data release of the Gaia mission (Gaia~DR3; \citealt{GaiaCollaboration2023}). Gaia provides homogeneous measurements of stellar positions, parallaxes, proper motions, and broad-band photometry, making it particularly suitable for identifying cluster members and characterizing open-cluster populations. For Berkeley~31, we queried the Gaia~DR3 archive for all sources located within a circular region of $20$~arcmin centered on the cluster. The retrieved catalog includes the equatorial coordinates ($\alpha$, $\delta$), trigonometric parallax ($\varpi$), proper motions ($\mu_{\alpha}\cos\delta$, $\mu_{\delta}$), and $G$, $G_{\rm BP}$, and $G_{\rm RP}$ photometric measurements. Appendix~\ref{app:data_quality} provides the adopted quality criteria and details of the Gaia data selection.

\subsection{Gaia-ESO Spectroscopy}
\label{subsec:spectroscopy_data}

Spectroscopic information was obtained from the fifth data release of the Gaia-ESO Survey \citep[GES~DR5.1;][]{Hourihane2023}, corresponding to the survey's sixth internal data release (iDR6). The GES observations were carried out with the FLAMES multi-object spectrograph at the ESO Very Large Telescope \citep{Pasquini2002}, providing homogeneous spectroscopic measurements for stars observed in Galactic stellar populations and OCs. The DR5.1 catalog contains stellar atmospheric parameters and elemental abundances derived through the survey's analysis framework and subsequently homogenized onto a common scale \citep{Hourihane2023}. We cross-matched the spectroscopic catalog with the Gaia~DR3 sources identified as probable Berkeley~31 members in our membership analysis, using a positional tolerance of $1\arcsec$. The initial cross-match resulted in 60 common sources.

Starting from 60 spectroscopic measurements associated with the probable cluster members, we imposed a signal-to-noise ratio (SNR) criterion of $\mathrm{SNR}\geq15$, resulting in a final sample of 40 stars. Four sources in the initial sample were flagged as NIA in the spectroscopic quality information; however, all four had $\mathrm{SNR}<15$ and were therefore excluded from the final sample by the adopted signal-to-noise threshold. The resulting sample provides a statistically meaningful basis for the chemical characterization of Berkeley~31 while limiting the impact of low-quality spectra on the derived abundances. The GES data provide measurements of the effective temperature ($T_{\rm eff}$), surface gravity ($\log g$), iron abundance ($[\mathrm{Fe/H}]$), and individual elemental abundances.

\subsection{Swift/UVOT Ultraviolet Photometry}

We obtained ultraviolet (UV) photometry, which is particularly useful for identifying hot components associated with BSS systems, from the Ultraviolet/Optical Telescope (UVOT) aboard the Neil Gehrels Swift Observatory \citep{gehrels2004swift}. The UVOT provides a $17\arcmin\times17\arcmin$ field of view and UV imaging suitable for detecting hot stellar components. For Berkeley~31, we used the UVOT photometry available in the $UVW1$, $UVM2$, and $UVW2$ filters, with effective wavelengths of approximately 2600, 2246, and 2086~\AA, respectively \citep{Breeveld2011}.

\begin{figure}[ht]
    \centering
    \includegraphics[width=1\linewidth]{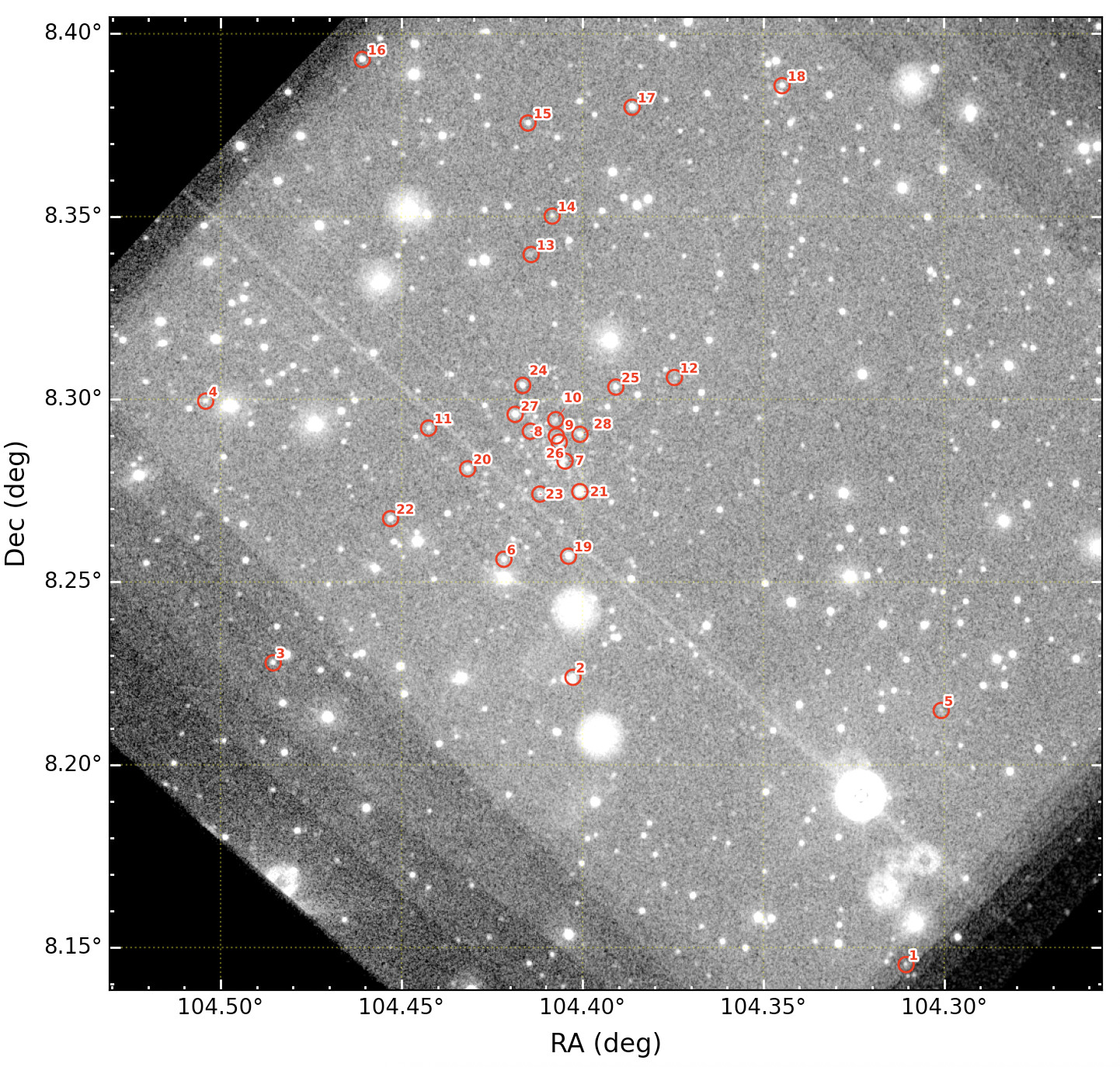}
    \caption{Combined \textit{Swift}/UVOT image of the Berkeley~31 field, constructed by overlaying observations obtained at different epochs. 28 BSSs with available \textit{Swift}/UVOT photometry are marked.}
    \label{fig:B31_UVOT_QC}
\end{figure}

To illustrate the spatial distribution of the UV-detected BSSs, we combined the \textit{Swift}/UVOT images\footnote{\url{https://www.swift.ac.uk/swift_portal/}} obtained at different epochs and overlaid them to show the common field of view. The 28 BSSs with available \textit{Swift}/UVOT photometry are marked in the combined image shown in Figure~\ref{fig:B31_UVOT_QC}.

Rather than reducing the UVOT images independently, we used the calibrated UVOT flux measurements provided by \citet{Siegel2019}. These measurements are based on the Level~3 point-source catalogs available through the \textit{Swift}/UVOT archive. The cataloged UV measurements were cross-matched with the Gaia sources selected as cluster members and were used to investigate the ultraviolet properties of the BSS candidates. The observational details and the UVOT measurements used in this study are provided in Appendix~\ref{app:swift_data}.

\subsection{Zwicky Transient Facility}

We used archival time-series photometric data from the Zwicky Transient Facility (ZTF) in the $i$ band, obtained through the ZTF IRSA archive. We accessed the ZTF data using the IRSA ZTF Time Series Tool and retrieved the time-series magnitudes and associated uncertainties for our target stars. ZTF is a wide-field optical survey conducted with the Samuel Oschin 48-inch Telescope at Palomar Observatory \citep{bellm2019zwicky}, and \ citet {masci2019zwicky} describe its data processing and archival products.

\subsection{Additional Multi-wavelength Photometry}

To extend the wavelength coverage of the SEDs beyond the Gaia and \textit{Swift}/UVOT measurements, we included archival photometry from several optical and infrared surveys. Optical photometric measurements were obtained from the Pan-STARRS1 Data Release 2 \citep[DR2;][]{2020ApJS..251....7F} and the SkyMapper Southern Survey Data Release 4 \citep[DR4;][]{2024PASA...41...61O}. For the infrared part of the SEDs, we used near-infrared photometry in the $J$, $H$, and $K_{\rm s}$ bands from the Two Micron All Sky Survey \citep[2MASS;][]{2006AJ....131.1163S}, together with mid-infrared measurements from the AllWISE data release of the Wide-field Infrared Survey Explorer \citep[WISE;][]{2014yCat.2328....0C}. We used the available measurements from these surveys as additional flux points in the SED analysis. The photometric data were retrieved through the Virtual Observatory SED Analyzer \citep[VOSA;][]{Bayo2008}.

\section{Membership Analysis}
\label{sec:membership}
\begin{figure*}
    \centering
    \includegraphics[width=0.95\linewidth]{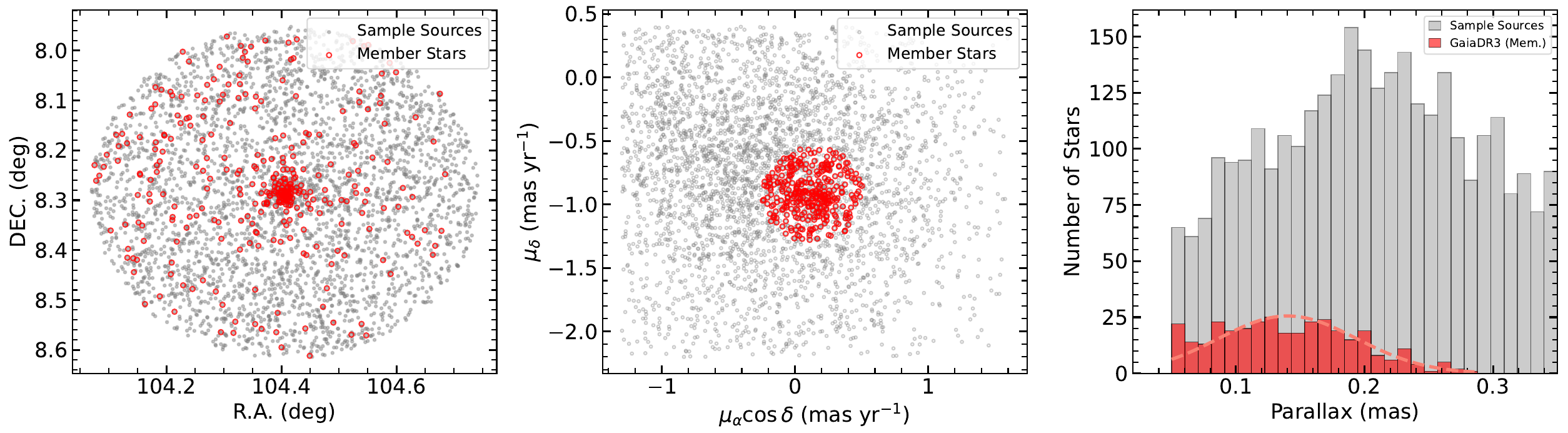}
   \caption{Spatial, kinematic, and astrometric distributions of the Gaia~DR3 sources toward Berkeley~31. Left: spatial distribution of the sources in equatorial coordinates ($\alpha$, $\delta$), with probable cluster members highlighted in red. Middle: proper-motion vector-point diagram showing the concentration of the selected members around the cluster proper-motion locus. Right: trigonometric parallax distribution of the full sample (grey) and the probable cluster members (red), with the Gaussian fit to the member-star parallax distribution shown by the red dashed curve. The probable members are selected from the GMM-based membership analysis with $P\geq0.6$.}     
    \label{fig:spatial_kinematic_dist}
\end{figure*}

Reliable identification of cluster members is essential for the photometric, chemical, and dynamical analysis of Berkeley~31. Although Gaia-based membership catalogs exist, we derived an independent Gaia~DR3 membership sample to ensure a consistent selection procedure across this study. To reliably identify cluster members, we employed the method of \citet{agarwal2021ml}, which has also been adopted in our previous studies \citep{belwal2025unveiling, belwal2026time}. Membership probabilities were determined in the three-dimensional astrometric space defined by the proper-motion components, $\mu_{\alpha}\cos\delta$ and $\mu_{\delta}$, and the trigonometric parallax, $\varpi$. We used a Gaussian Mixture Model (GMM; \citealt{dempster1977maximum,mclachlan2000finite}) to separate the cluster and field populations and assigned the probability of belonging to the cluster component as the membership probability ($P$).

Before fitting the mixture model, we applied the astrometric and photometric quality criteria described in Appendix~\ref{app:data_quality} to the Gaia DR3 sources within a 20 arcmin search region. Sources were required to have a full five-parameter astrometric solution and reliable Gaia photometry. We also applied the adopted limits on parallax and proper-motion uncertainties, together with a renormalized unit weight error $\mathrm{RUWE}\leq 1.4$, to exclude sources with potentially unreliable astrometric measurements. RUWE provides a measure of the goodness of fit of the Gaia astrometric solution, and the adopted threshold helps reduce the contribution of sources with poorly fitted astrometric measurements to the membership analysis \citep{Lindegren2018, Lindegren2021}.

The approximate location of the cluster population in $(\mu_{\alpha}\cos\delta,\mu_{\delta},\varpi)$ space was first identified using a k-nearest neighbors (kNN) approach \citep{cover1967nearest}. This method examines the local distribution of stars in astrometric parameter space and was used to identify the concentration associated with the cluster before fitting the mixture model. The three astrometric quantities were then standardized and used to initialize a two-component GMM fitted with the Expectation--Maximization algorithm. One component represents the concentrated cluster distribution, and the other the broader field population. We adopted the posterior probability of the cluster component as $P$ for each source.

Stars with $P\geq0.6$ were selected as probable members of Berkeley~31. We adopted this threshold to balance sample completeness against a clear concentration around the cluster locus in astrometric space. We also tested small variations around this limit and found that the main cluster properties were not strongly affected by the adopted threshold. The final catalog contains 329 stars with $P\geq0.6$. The selected members form a distinct concentration in proper-motion and parallax space, and their spatial distribution is consistent with a centrally concentrated stellar system (Figure~\ref{fig:spatial_kinematic_dist}).

As an independent check, we compared our membership sample with the Gaia-based catalog of \citet{Hunt2023}, probable members of Berkeley~31, with membership probabilities ranging from 0.57 to 1. We did not adopt this catalog directly because its membership criteria differ from ours. Instead, it was used as an external reference to assess the consistency of our independent selection. All 201 stars listed by \citet{Hunt2023} are recovered in our catalog, while our analysis also identifies additional probable members. A detailed comparison between the two membership samples is presented in Section~\ref{sec:comp_membership}.

We then determined the cluster center from the spatial distribution of the selected members. Independent one-dimensional stellar-density profiles were constructed along right ascension and declination and fitted with Gaussian functions. The fitted maxima were adopted as the corresponding cluster-center coordinates. We obtain $\alpha = 06^{\mathrm h}57^{\mathrm m}37^{\mathrm s}.45$ and $\delta = +08^\circ17'13''.70$, corresponding to Galactic coordinates $l = 206^\circ.240757$ and $b = 5^\circ.134826$. We computed the mean astrometric properties of the selected members using their Gaia~DR3 measurements. The mean proper motions and trigonometric parallax are listed in Table~\ref{tab:summary}. These values define the characteristic astrometric locus of Berkeley~31 and are used in the subsequent photometric and dynamic orbital analyzes.

\section{Fundamental Properties}
\label{sec:fundamental_properties}

Following the Gaia~DR3 membership determination, we characterize the fundamental physical properties of Berkeley~31 using the resulting member sample. We first determine the cluster distance and line-of-sight extinction, and independently characterize its chemical composition using spectroscopic abundances of iron and $\alpha$ elements. We then compare the stellar population with theoretical isochrones to constrain the cluster age and related photometric properties. Finally, we estimate the fraction of unresolved binary systems among the cluster members. Together, these analyses provide a self-consistent set of photometric and chemical constraints on Berkeley~31, which underpin our discussion of its stellar populations, dynamical state, and evolutionary history in the following sections.

\subsection{Distance Estimation and Interstellar Reddening}
\label{sec:distance_reddening}

We estimated the distance to Berkeley~31 using the Gaia~DR3 parallaxes of its cluster members. Following \citet{Lindegren2021}, we applied a parallax zero-point correction of $-0.017$~mas, such that the corrected parallax of each star is given by $\varpi_{\rm corr}=\varpi_{\rm Gaia}+0.017$~mas. Instead of converting individual parallaxes to distances before averaging, we computed the mean parallax directly in parallax space. This approach avoids inverting noisy individual parallaxes and helps reduce biases associated with fractional parallax uncertainties, including the Lutz--Kelker effect \citep[e.g.,][]{Lutz1973,Smith2003,Eker2009}. We excluded BSSs from this calculation because they represent a distinct evolutionary population and are not part of the normal cluster sequence used to determine the cluster distance. We then obtained the distance from the corrected mean parallax as $d=1000/\langle\varpi_{\rm corr}\rangle$, where $d$ is in pc and $\langle\varpi_{\rm corr}\rangle$ is in mas.

\begin{figure}
\centering
\includegraphics[width=0.94\linewidth]{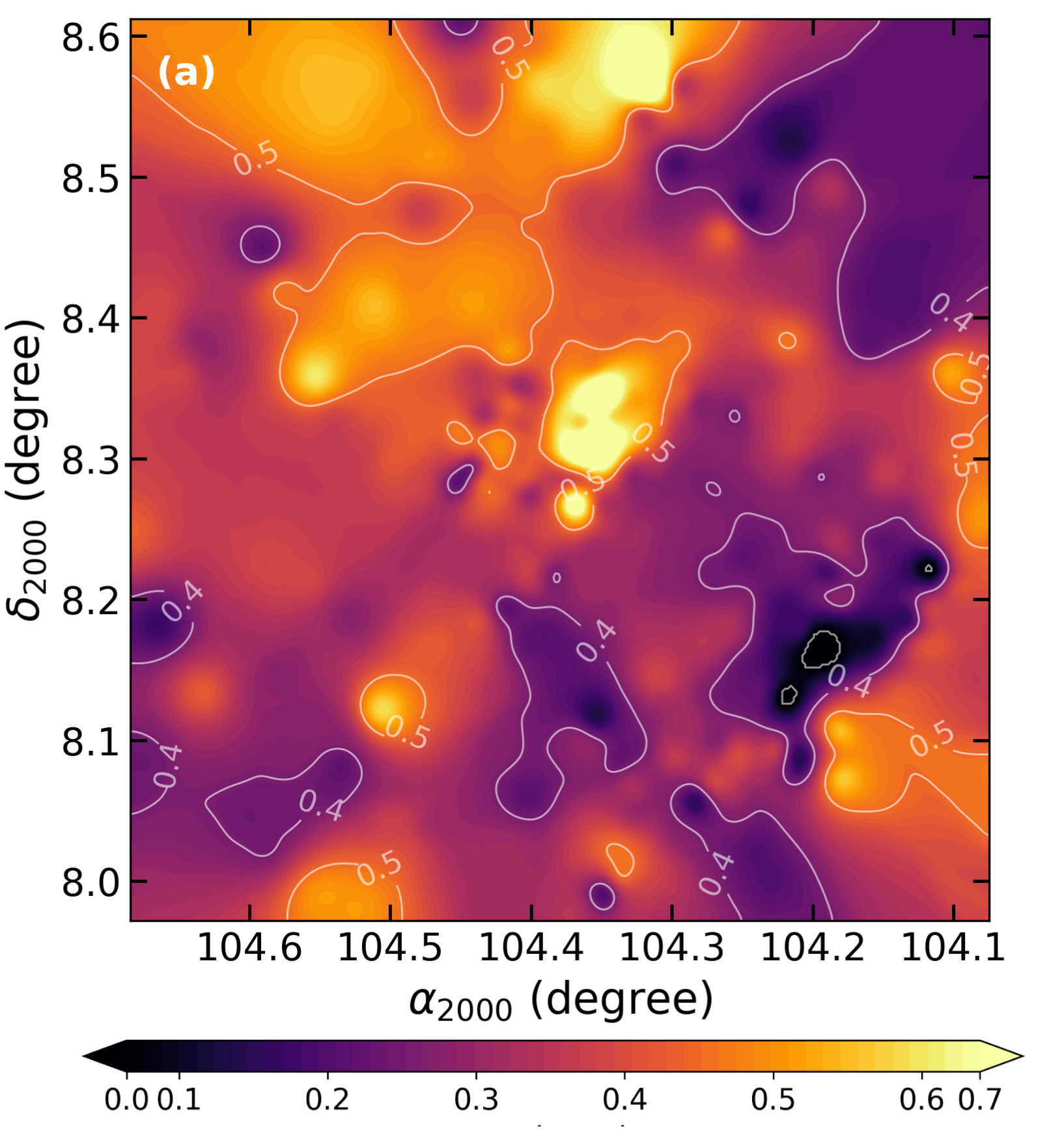}
\includegraphics[width=0.99\linewidth]{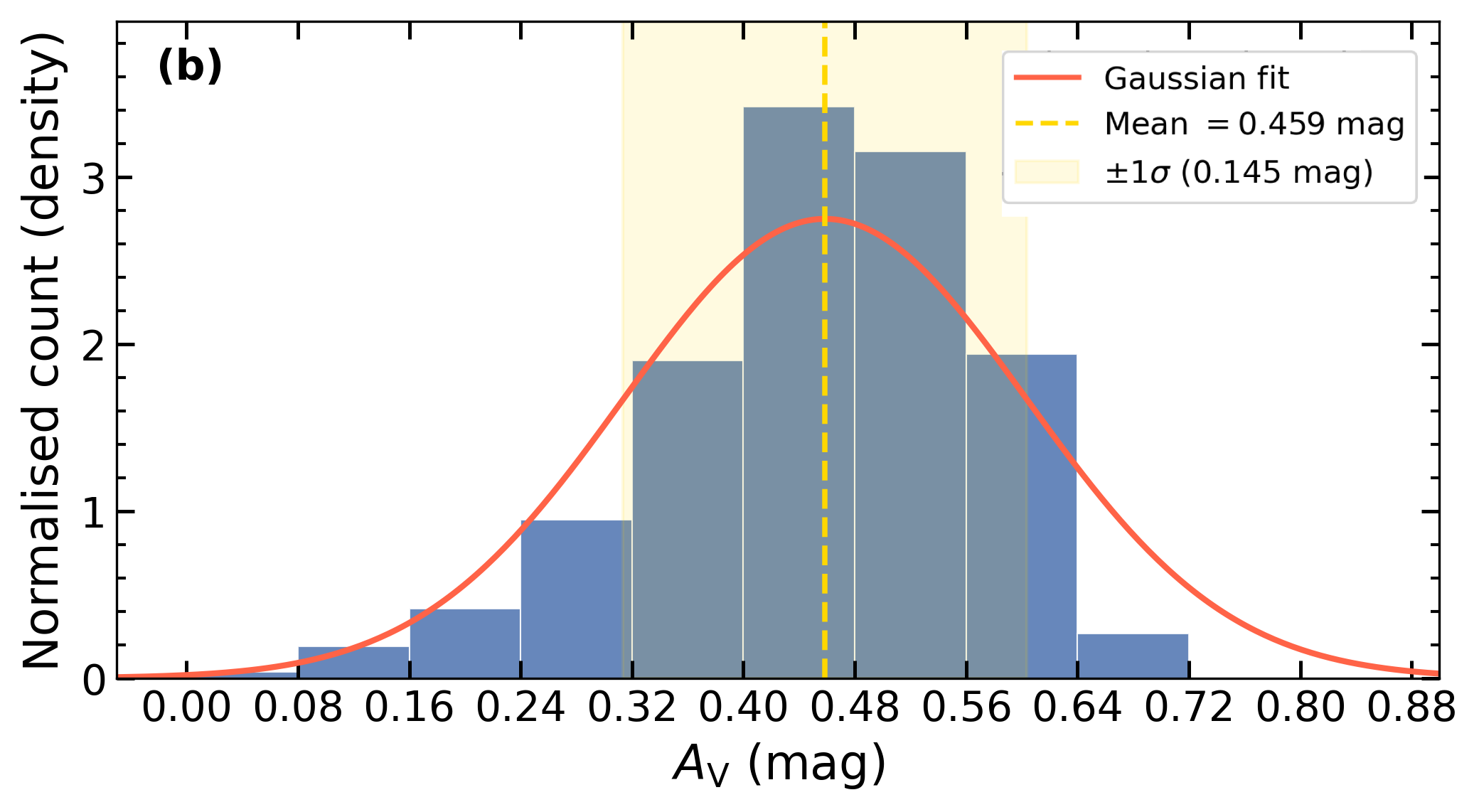}
\caption{Interstellar extinction toward Berkeley~31 derived from the three-dimensional {\sc Bayestar19} dust map \citep{Green2019}. (a) Spatial distribution of the visual extinction $A_{\rm V}$ across the cluster field. The color scale and contours trace extinction variation along the line of sight to the member stars. (b) Distribution of the individual $A_{\rm V}$ estimates. The solid curve shows the Gaussian fit to the distribution, the dashed line marks the mean, and the shaded region indicates the $1\sigma$ interval.}
\label{fig:AV}
\end{figure}

As shown in the CMD in Figure~\ref{fig:B31_CMD_membership} (Section~\ref{sec:comp_membership}), MSTO region starts at approximately $G=17.1$ mag and extends to about $G=17.4$ mag. We classified stars with $G<17.1$ mag as giant stars and those with $G>17.4$ mag as main-sequence (MS) stars, while stars in the $17.1<G ~{\rm (mag)} <17.4$  interval were considered MSTO stars. We excluded BSS candidates from these populations. We calculated the mean corrected parallax separately for the giant, MSTO, and MS populations to check their consistency. Figure~\ref{fig:AV} shows the resulting distributions. The 36 giant members have $\langle\varpi_{\rm corr}\rangle=0.170\pm0.006$~mas, corresponding to $d=5882\pm 208$~pc. For the 40 MSTO stars, we obtain $\langle\varpi_{\rm corr}\rangle=0.163\pm0.015$~mas and $d=6135\pm 565$~pc, while the 189 MS members give $\langle\varpi_{\rm corr}\rangle=0.150\pm0.012$~mas and $d=6667\pm 533$~pc. For all 265 non-BSS members, the mean corrected parallax is $\langle\varpi_{\rm corr}\rangle=0.166\pm0.005$~mas, giving $d=6024\pm 181 $~pc. The quoted distance uncertainty of $181$~pc was derived from the uncertainty in the mean parallax and does not include the systematic uncertainty associated with the parallax zero-point correction.

As an additional consistency check, we repeated the analysis using only stars in the magnitude range where the mean parallax uncertainty is approximately $0.017$~mas. The resulting distance, $d=6024\pm 181 $~pc, is consistent with that derived from the full sample of 265 non-BSS members, supporting the robustness of the adopted cluster distance.

\begin{figure}
\centering
\includegraphics[width=0.98\linewidth]{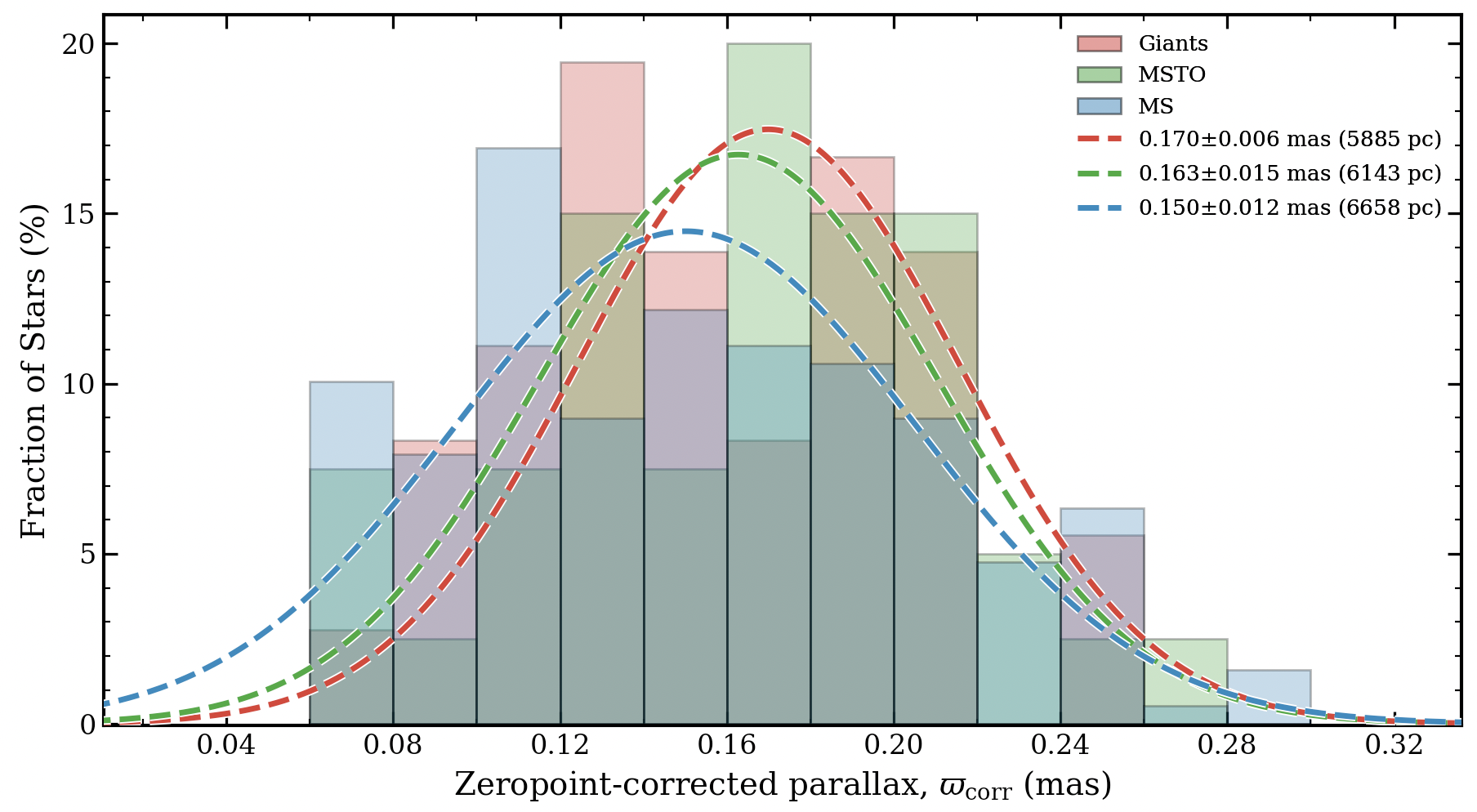}
\caption{Distributions of the zero-point-corrected Gaia parallaxes ($\varpi_{\rm corr}$) for the giant (red), MSTO (green), and MS (blue) member populations of Berkeley~31. The dashed curves show the corresponding Gaussian fits. The legend gives the fitted mean parallaxes and distances.}
\label{fig}
\end{figure}

The MS stars have a somewhat smaller mean parallax than the giants, but this difference is not statistically significant and may partly result from the larger uncertainties of the fainter stars. We therefore do not interpret the differences between the evolutionary groups as evidence for a physical distance variation within the cluster. The population estimates are broadly consistent with a distance of about 6~kpc, and we adopt $d=6024\pm 181$~pc from the full non-BSS sample. At this distance, Gaia parallax uncertainties are particularly relevant for the faint MS stars. We therefore use the zero-point-corrected mean Gaia trigonometric parallax as the basis for the adopted distance rather than relying on individual or Bayesian distance estimates \citep[e.g.,][]{BailerJones2021}.

We investigated the interstellar reddening toward Berkeley~31 using the three-dimensional {\sc Bayestar19} dust map of \citet{Green2019}. For each cluster member, we evaluated the reddening at its sky position using the adopted cluster distance of $6024\pm181$~pc. This approach also allows us to assess spatial variations in foreground extinction across the cluster field.

The resulting extinction distribution and spatial variation are shown in Figure~\ref{fig:AV}. The extinction map (Figure~\ref{fig:AV}a) shows a structured but moderate dust distribution, with $A_{\rm V}$ values ranging from nearly zero to about 0.72~mag. Regions of higher extinction are interspersed with areas of lower extinction, indicating that the reddening is not completely uniform across the field. The overall extinction level remains moderate compared with strongly obscured Galactic-plane clusters.

The corresponding $A_{\rm V}$ distribution (Figure~\ref{fig:AV}b) is reasonably described by a single Gaussian, centered at $A_{\rm V}=0.459\pm0.145$~mag. The quoted uncertainty of $\pm0.145$~mag represents the uncertainty in the adopted mean extinction, rather than the star-to-star scatter in extinction. The lack of a pronounced secondary component suggests that the variations are mainly continuous rather than associated with distinct foreground dust populations. Assuming $R_{\rm V}=3.1$, the corresponding mean color excess is $\langle E(B-V)\rangle=0.148\pm0.047$~mag. The star-to-star variation indicates differential extinction across the field and should be considered in the photometric analysis.

\subsection{Metallicity and Chemical Abundance Pattern}\label{sec:metallicity}
\begin{figure}
\centering
\includegraphics[width=0.98\linewidth]{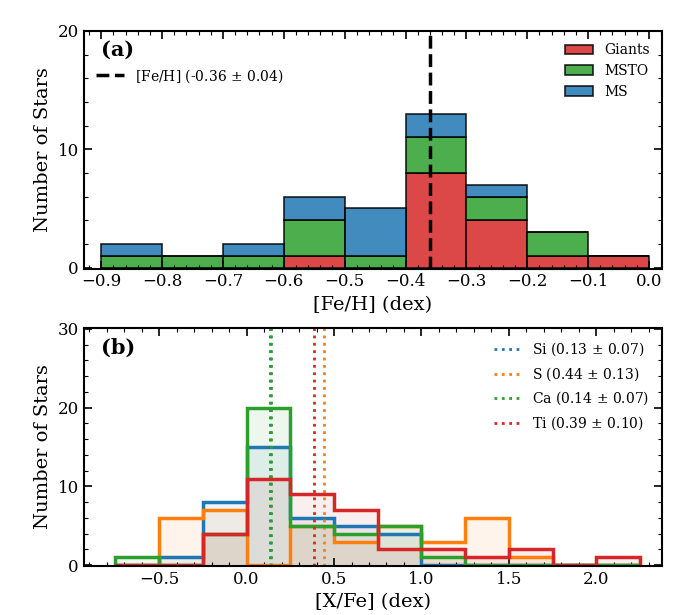}
\caption{Chemical abundance distributions of Berkeley~31 member stars satisfying the $\mathrm{SNR}\geq15$ criterion. Panel (a) shows the [Fe/H] distribution, with different colors indicating the evolutionary stage of the stars. The vertical dashed line marks the median iron abundance, $\mathrm{[Fe/H]}=-0.36\pm0.04$~dex. Panel (b) presents the distributions of the available $\alpha$-element abundance ratios, [Si/Fe], [S/Fe], [Ca/Fe], and [Ti/Fe]. The vertical dotted lines indicate the corresponding median abundance values, with the associated uncertainties representing the median uncertainty.}
\label{fig:element_histo}
\end{figure}

The [Fe/H] distribution of the selected stars is shown in Figure~\ref{fig:element_histo}a. The individual measurements cover approximately $-0.9\lesssim\mathrm{[Fe/H]}\lesssim-0.1$~dex, with most stars clustered around $\mathrm{[Fe/H]}\simeq-0.4$~dex. The median iron abundance is $\mathrm{[Fe/H]}=-0.36\pm0.04$~dex, which we adopt as the representative metallicity of Berkeley~31. The uncertainty in this value was estimated by accounting for the individual spectroscopic abundance uncertainties of the selected member stars. Thus, the quoted uncertainty reflects the uncertainty associated with the combined abundance measurements; systematic effects common to the spectroscopic analysis are not independently quantified here. The observed spread is larger than expected for a cluster and is likely related to individual abundance uncertainties, differences in evolutionary stage, and spectral quality rather than an intrinsic metallicity spread.

\begin{figure*}
\centering
\includegraphics[width=1\linewidth]{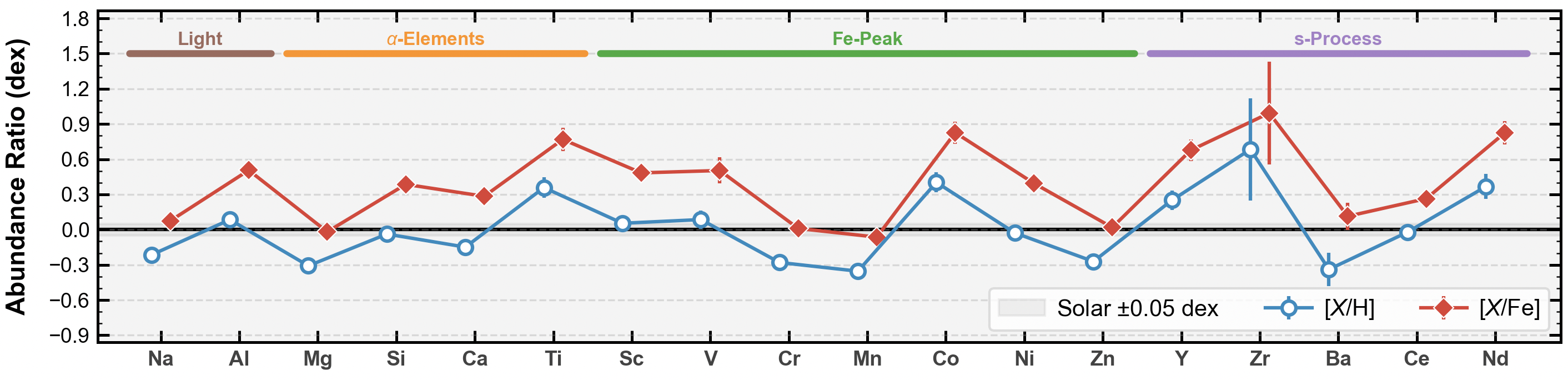}
\caption{Elemental abundance pattern of Berkeley~31 derived from the available spectroscopic measurements of cluster members. The blue circles show [X/H] abundance ratios, while the red diamonds represent [X/Fe] abundance ratios, both in dex. The elements are grouped into light, $\alpha$-, Fe-peak, and s-process species by nucleosynthetic classification. The horizontal dashed line indicates the solar abundance ratio.}
\label{fig:abundance_pattern}
\end{figure*}

The available $\alpha$-element measurements include Si, S, Ca, and Ti, with measurements for 39, 36, 40, and 39 stars, respectively (Figure~\ref{fig:element_histo}b). Their median abundance ratios are [Si/Fe] $=0.13\pm0.07$, [S/Fe] $=0.44\pm0.13$, [Ca/Fe] $=0.14\pm0.07$, and [Ti/Fe] $=0.39\pm0.10$~dex. All four elements have positive median [X/Fe] values, with Ti showing the largest enhancement and S the largest uncertainty. The abundance pattern of Berkeley~31, including the [X/H] and [X/Fe] ratios for the available elements, is shown in Figure~\ref{fig:abundance_pattern}.

We calculated the mean $\alpha$-element abundance by combining the median [X/Fe] ratios of Si, S, Ca, and Ti using inverse-variance weighting. For each element, the weight was defined as $w_i=1/\sigma_i^2$, where $\sigma_i$ is the uncertainty associated with its median abundance ratio. The weighted mean and its formal uncertainty were calculated as:
\begin{equation}
\langle[\alpha/\mathrm{Fe}]\rangle =
\frac{\sum_i w_i [X_i/\mathrm{Fe}]}{\sum_i w_i},
\qquad
\sigma_{\langle[\alpha/\mathrm{Fe}]\rangle}
=\left(\sum_i w_i\right)^{-1/2}
\end{equation}
Combining the four measurements in this way gives $\langle[\alpha/\mathrm{Fe}]\rangle=0.21\pm0.04$~dex. For visual context, we include a $\pm0.05$~dex shaded band in Figure~\ref{fig:abundance_pattern} to indicate the typical systematic accuracy floor of standard 1D LTE stellar abundance analyses \citep[e.g.,][]{Jofre2019}.

For the stellar evolution models, we converted the median iron abundance to the heavy-element mass fraction $Z$ using the calibrated relation \citep[e.g.,][]{Elsanhoury2025, Alzhrani2025b, Alzhrani2025, Cinar2025, Cinar2026, Bisht2026b, Bisht2026c}:
\begin{equation}
Z_{\rm x} = \frac{Z}{0.7515 - 2.78 \times Z},
\end{equation}
\begin{equation}
\mathrm{[Fe/H]} =
\log(Z_{\rm x}) -
\log\left(
\frac{Z_{\odot}}
{1-0.248-2.78\times Z_{\odot}}
\right),
\end{equation}
where $Z_{\odot}=0.0152$. For $\mathrm{[Fe/H]}=-0.36\pm0.04$~dex, this gives $Z=0.00685\pm0.00061$, which we use in the subsequent isochrone analysis.

\subsection{Isochrone Fitting}

\begin{figure}
\centering
\includegraphics[width=0.95\linewidth]{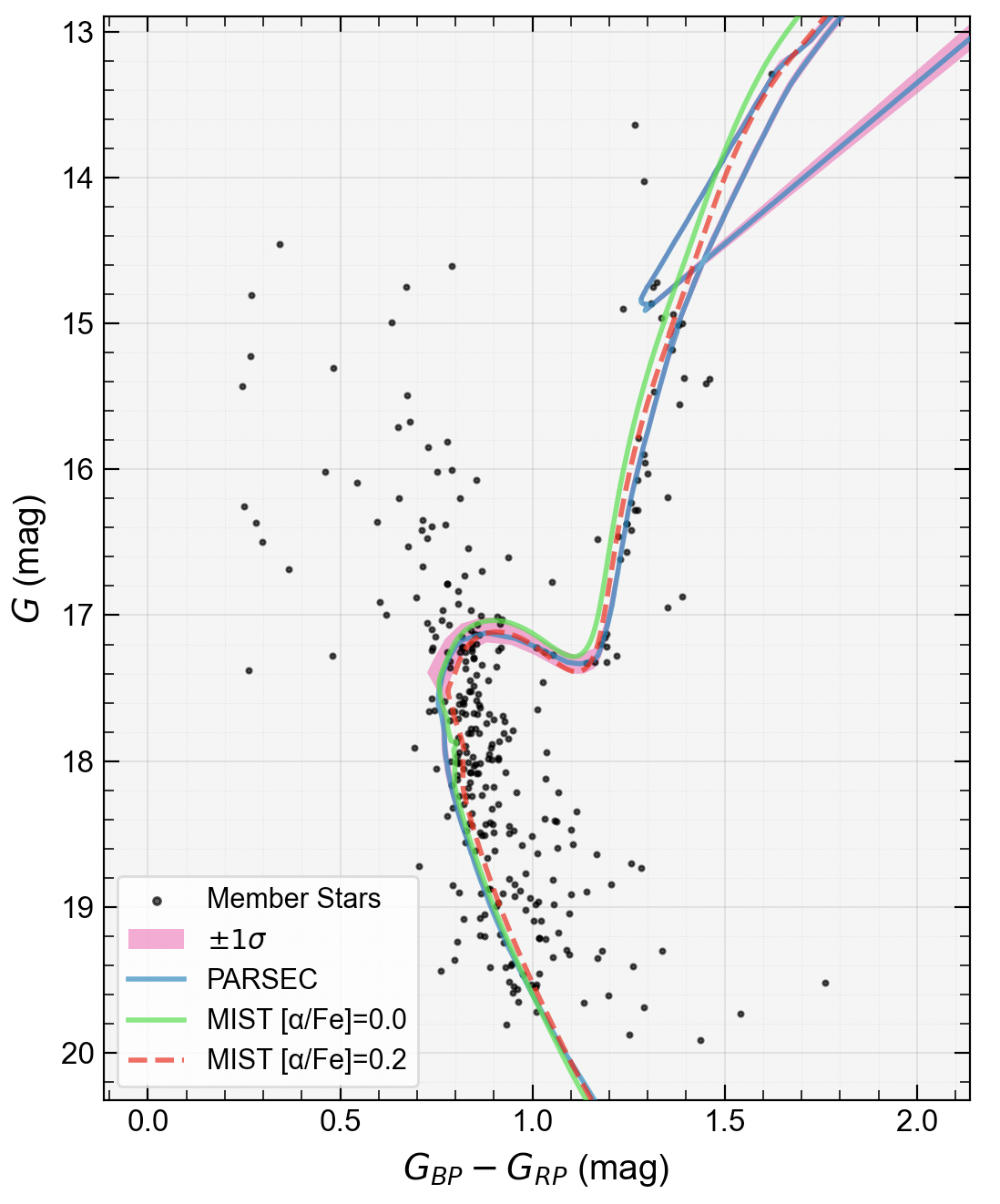}
\caption{CMD of the high-probability member stars of Berkeley~31. The black points denote the astrometrically selected cluster members. The solid blue curve corresponds to the PARSEC isochrone, while the solid green and dashed red curves show the MIST models computed for $[\alpha/{\rm Fe}]=0.0$ and $0.2$~dex, respectively. The pink shaded region represents the $\pm1\sigma$ age uncertainty of $0.2$~Gyr around the best-fit isochrone.}
\label{fig:CMD_B31}
\end{figure}

We further examined the evolutionary state of Berkeley~31 by comparing the CMD of its probable members with theoretical stellar isochrones. Figure~\ref{fig:CMD_B31} shows the $G$ versus $G_{\rm BP}-G_{\rm RP}$ CMD constructed from the stars identified as cluster members in our astrometric analysis, with a well-defined MS, MSTO region, and sequence of evolved stars toward brighter magnitudes. For the isochrone comparison, distance, extinction, and chemical composition were fixed to the independently determined cluster parameters listed in Table~\ref{tab:summary}, rather than treated as free parameters. The resulting comparison provides an estimate of the cluster age based on the observed CMD morphology.

We considered two evolutionary models, using PARSEC models generated with CMD~3.9 \citep{Bressan2012} and MIST isochrones from version~2.5 \citep{Choi2016, Dotter2016}. For the MIST models, we adopted two $\alpha$-element abundance levels, $[\alpha/{\rm Fe}]=0.0$ and $0.2$~dex, with the latter close to the spectroscopically measured value for Berkeley~31. This allows us to examine the effect of a moderate $\alpha$-element enhancement on the isochrone comparison. The age uncertainty of $0.2$ Gyr was estimated by shifting the isochrone to account for the observed spread around the MSTO and evolved-star regions, while avoiding the lower MS, which is more affected by unresolved binaries, and the BSS population.

As shown in Figure~\ref{fig:CMD_B31}, the three model sequences follow the observed MS closely over the magnitude range where the member sample is most densely populated. Around the MSTO, small offsets between the PARSEC and MIST predictions become apparent, while the differences are more noticeable along the brighter evolved sequence. Nevertheless, the different model sets consistently reproduce the principal features of the CMD. The two MIST solutions with $[\alpha/{\rm Fe}]=0.0$ and $0.2$~dex remain relatively close, indicating that the moderate $\alpha$-enhancement adopted here does not substantially alter the overall age solution.

\subsection{Chemical Clock Ages}\label{sec:chemical_age}

As an independent assessment of the isochrone-based age, we derived chemical ages for Berkeley~31 using abundance-based stellar clocks. We considered the multivariate chemical-clock relations of \citet{Viscasillas2022}, which express stellar age as a function of an abundance ratio, Galactocentric distance, and metallicity. Given the location of Berkeley~31 at $R_{\rm gc}\simeq 13.85\pm0.20$~kpc, we adopted the outer-disk calibration of these relations. Four Ba-based abundance ratios, [Ba/Al], [Ba/Mg], [Ba/Si], and [Ba/Ca], were considered. Before calculating the ages, the individual abundances were quality-controlled following the procedure of \citet{Viscasillas2022}; measurements with uncertainties of $\geq0.1$~dex were excluded, and outlying abundance measurements were removed using the interquartile-range criterion. We placed the remaining abundances on the appropriate reference scale according to each star's evolutionary status. We then evaluated the calibrated relations separately for each star and each available abundance ratio.

The mean ages obtained from the four Ba-based clocks are $3.29\pm1.43$, $3.96\pm0.59$, $3.44\pm0.83$, and $5.02\pm1.37$~Gyr for [Ba/Al], [Ba/Mg], [Ba/Si], and [Ba/Ca], respectively. The somewhat higher age obtained from [Ba/Ca] could be related to the different nucleosynthetic origins of Ba and Ca, as Ba is mainly associated with $s$-process enrichment in Asymptotic Giant Branch (AGB) stars, whereas Ca is an $\alpha$-element \citep{Cseh2018}. These independent estimates broadly cover the same age range, despite differences in sensitivity and sample size across the individual relations. To obtain a combined chemical-clock estimate, we considered the four Ba-based ages on a star-by-star basis. We excluded individual age estimates that yielded negative, non-physical ages and averaged the remaining valid clock ages for each star. We then obtained the final chemical age as the mean of the resulting stellar ages for the 11 stars with at least one valid Ba-based chemical-clock estimate, yielding $3.72\pm1.42$~Gyr.

\begin{table}[!ht]
\centering
\scriptsize
\renewcommand{\arraystretch}{0.95}
\setlength{\tabcolsep}{0.9pt}
\caption{Derived physical and kinematic parameters of Berkeley~31.}
\label{tab:summary}
\begin{tabular}{lcr}
\hline\hline
Parameter & Symbol & This study \\
\hline
\multicolumn{3}{c}{\textit{Astrometric Parameters}} \\
\hline
Right ascension                 & $\alpha$ (degree)  &104.406044 \\
Decl.                     & $\delta$ (degree)  & 8.287138\\
Galactic longitude              & $l$ (degree)         & 206.241 \\
Galactic latitude               & $b$ (degree)         & $+$5.135 \\
Number of members      & $N_{\rm mem}$        & 329 \\
Mean proper motion (RA)         & $\langle\mu_{\alpha}\cos\delta\rangle$ (mas yr$^{-1}$) & 0.110$\pm$0.215 \\
Mean proper motion (Dec)        & $\langle\mu_{\delta}\rangle$ (mas yr$^{-1}$)        & -0.920$\pm$0.186 \\
Mean trigonometric parallax     & $\langle\varpi\rangle$ (mas)                & 0.166$\pm$0.005 \\
Mean parallax distance          & $\langle d_{\varpi}\rangle$ (pc)             & 6024$\pm$181 \\
\hline
\multicolumn{3}{c}{\textit{Fundamental Parameters}} \\
\hline
True distance modulus           & $(m-M)_0$ (mag)                   & 13.895$\pm$0.062 \\
Apparent distance modulus       & $(m-M)$ (mag)                     & 14.279$\pm$0.136 \\
$V$-band extinction              & $\langle A_{\rm V} \rangle$ (mag) & 0.459$\pm$0.145 \\
Color excess ($UBV$)             & $E(B-V)$ (mag)                    & 0.148$\pm$0.047 \\
Color excess (Gaia)              & $E(G_{\rm BP}-G_{\rm RP})$ (mag)  & 0.209$\pm$0.066 \\
$G$-band extinction               & $\langle A_{\rm G}\rangle$ (mag) & 0.384$\pm$0.121 \\
Iron abundance                  & $[\rm{Fe/H}]$ (dex)                 & -0.36$\pm$0.04 \\
Metallicity                     & $Z$                                 & 0.00685$\pm$0.00061 \\
$\alpha$-element abundance      & $\langle[\alpha/\mathrm{Fe}]\rangle$ (dex) & 0.21$\pm$0.04 \\
Age (isochrone)                 & $t_{\rm iso}$ (Gyr)                 &  3.7$\pm$0.2 \\
Age (chemical clocks)    & $t_{\rm chem}$ (Gyr) & 3.72$\pm$1.42 \\
Unresolved binary fraction      & $f_{\rm bin}$                       & 0.38$\pm$0.04  \\
\hline
\multicolumn{3}{c}{\textit{Galactic Orbital Parameters}} \\
\hline
Radial velocity                 & $\langle V_{\rm rad,OC}\rangle$ (km s$^{-1}$) & $57.89\pm0.08$  \\
Perigalactic radius             & $R_{\rm peri}$ (kpc)              & $11.08\pm0.37$  \\
Apogalactic radius              & $R_{\rm apo}$ (kpc)               & $13.87\pm0.18$  \\
Mean orbital radius             & $R_{\rm m}$ (kpc)                 & $12.48\pm0.21$  \\
Orbital eccentricity            & $e$                               & $0.112\pm0.010$  \\
Maximum vertical height         & $Z_{\rm max}$ (pc)                & $592\pm92$  \\
Present Galactocentric radius   & $R_{\rm GC}$ (kpc)                & $13.85\pm0.20$  \\
Guiding radius                  & $R_{\rm g}$ (kpc)                 & $12.29\pm0.29$  \\
Chemical birth radius           & $R_{\rm b}$ (kpc)                 & $11.15\pm0.50$  \\
Orbital period                  & $T_{\rm p}$ (Myr)                & $365.5\pm9.0$  \\
\hline
\multicolumn{3}{c}{\textit{Space Velocities}} \\
\hline
Velocity component & $U_{\rm LSR}$ (km s$^{-1}$) & 39.55$\pm$2.57  \\
Velocity component & $V_{\rm LSR}$ (km s$^{-1}$) & -21.06$\pm$5.02  \\
Velocity component & $W_{\rm LSR}$ (km s$^{-1}$) & 2.78$\pm$5.87  \\
Space velocity & $S_{\rm LSR}$ (km s$^{-1}$) & 44.89$\pm$8.13  \\
\hline
\hline
\end{tabular}
\end{table}

\section{Kinematic and Dynamical Properties}
\label{sec:dynamical_char}

\subsection{Radial Velocity Determination}
\label{subsec:radial_velocity}

\begin{figure}
    \centering
    \includegraphics[width=1\linewidth]{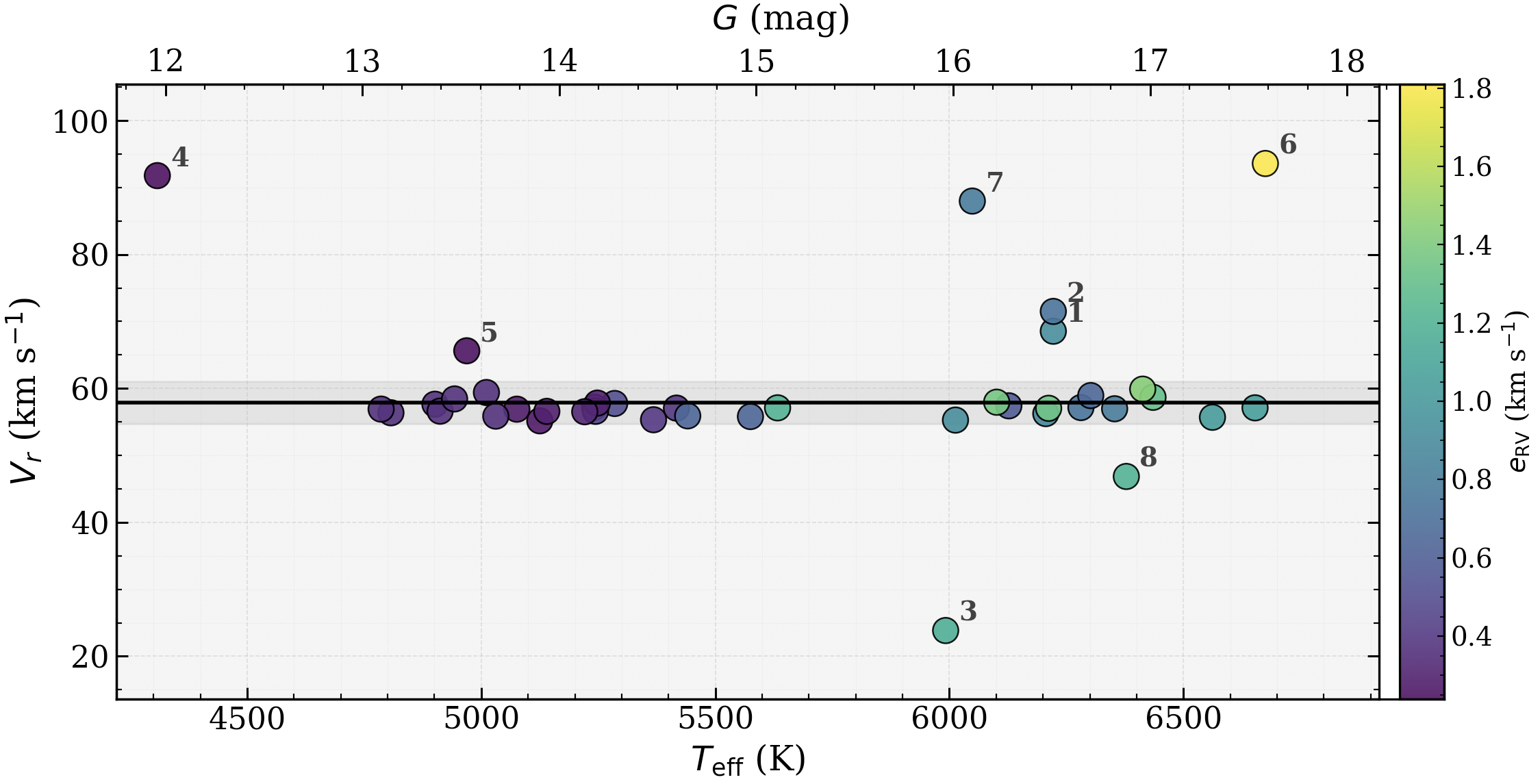}
    \caption{RVs of the Berkeley~31 members plotted against effective temperature. Individual points are color-coded according to their RV uncertainties. The horizontal solid line indicates the adopted systemic RV of the cluster, while the shaded band represents the corresponding $1\sigma$ interval. The eight sources outside the initial $1\sigma$ interval are labeled from 1 to 8, and their corresponding Gaia-ESO IDs are provided in the text.}
    \label{fig:radial_vel_b31}
\end{figure}

The systemic radial velocity (RV) of Berkeley~31 was determined from the spectroscopic measurements available in the Gaia-ESO Survey DR5.1 \citep{Hourihane2023}. We considered 40 cluster members with radial-velocity measurements and combined their velocities using inverse-variance weighting, following \citet{Soubiran2018} and \citet{Carrera2022}. Each measurement was assigned a weight of $w_i=1/\sigma_{v_{\rm rad},i}^{2}$. An initial weighted mean and velocity dispersion were calculated, and stars lying outside the initial $1\sigma$ interval were excluded from the final estimate. This left 32 stars for the determination of the systemic velocity. The numerical labels of stars in Figure~\ref{fig:radial_vel_b31} correspond to the following Gaia-ESO IDs:

\begin{quote}
\textbf{1}: \texttt{06573777+0817399}; \textbf{2}: \texttt{06573785+0818307}; \textbf{3}: \texttt{06581159+0816454}; \textbf{4}: \texttt{06580108+0820505};\\
\textbf{5}: \texttt{06570496+0816357}; \textbf{6}: \texttt{06570050+0819505}; \textbf{7}: \texttt{06565772+0819593}; \textbf{8}: \texttt{06571911+0820275}.
\end{quote}

The resulting systemic RV is $\langle V_{\rm rad,OC}\rangle = 57.89\pm0.08$~km~s$^{-1}$, with a velocity dispersion of $3.17$~km~s$^{-1}$. The individual RVs are shown as a function of effective temperature in Figure~\ref{fig:radial_vel_b31}. The sources excluded by the initial clipping are not interpreted as non-members, since deviations in RV may also arise from binary motion or measurement variations. Their possible variability and orbital properties will be examined separately using the available multi-epoch spectroscopic observations.

To investigate the radial-velocity deviations of the eight stars from the mean cluster RV, we searched the available spectroscopic measurements in different catalogs and checked the available epochs. Six of the eight stars have two GIRAFFE spectra at different epochs in the Gaia-ESO Survey archive, which provides high-resolution spectroscopic data for stellar kinematics and chemical characterization \citep{Gilmore2022, Randich2022}. We retrieved these spectra from the ESO archive and measured their RVs using \texttt{iSpec} \citep{Blanco-Cuaresma2014}, adopting the ATLAS9/Sun spectrum as the reference. The spectra were shifted relative to the reference spectrum, and the RV corresponding to the best match was determined independently for each epoch. The two epochs are separated by approximately two months, allowing us to check for possible RV variations.

The resulting RV measurements show different levels of variation among the six stars. Stars 2 and 7 show small changes between the two epochs, with $\Delta V_{\rm rad}$ of 1.17 and 2.13 km s$^{-1}$, respectively. In contrast, Stars 1, 3, 6, and 8 show larger changes of 15.75, 11.03, 68.35, and 6.32 km s$^{-1}$, respectively. The particularly large change observed for Star 6, from $21.94\pm0.75$ to $90.29\pm0.72$ km s$^{-1}$, indicates significant RV variability. Stars 1 and 3 also show substantial changes between the two epochs, while Star 8 shows a comparatively smaller variation.

For Stars 4 and 5, only a single GIRAFFE epoch is available in the ESO archive, so a multi-epoch comparison from the GIRAFFE spectra is not possible. We therefore compared their available measurements with Gaia DR3 and literature RVs. Star 4 has a Gaia DR3 RV of $91.73\pm0.42$ km s$^{-1}$ based on eight transits, consistent with the $92.09\pm0.24$ km s$^{-1}$ value reported by \citet{Tarricq2021}. Its RV chi-square $p$-value of 0.50 shows no significant RV variability. Star 5  has a Gaia DR3 RV of $67.97\pm3.09$ km s$^{-1}$ from 13 transits, also consistent with the $65.93\pm0.25$ km s$^{-1}$ value of \citet{Tarricq2021}. Since individual epoch RVs and RV variability indicators are not available for this star, possible RV variability cannot be excluded. RV variations alone do not necessarily mean that these stars are non-members of Berkeley 31. Large RV differences between epochs may instead arise from unresolved binarity or orbital motion, which can shift the measured RV away from the cluster’s systemic velocity.

\subsection{Orbital Integration}
\label{subsec:orbit}

We investigated the Galactic orbit of Berkeley~31 using the \texttt{galpy} package \citep{Bovy2015}. The orbit was integrated in the axisymmetric \textsc{MWPotential2014} Galactic potential, which includes the disk, bulge, and dark-matter halo components \citep{Bovy2013}. As in our analysis of Berkeley~31, we also repeated the calculation using the \textsc{McMillan2017} potential \citep{McMillan2017}. The two models gave consistent orbital parameters within the uncertainties. We therefore adopted \textsc{MWPotential2014} for the results presented here.

\begin{figure}
    \centering
    \includegraphics[width=0.9\linewidth]{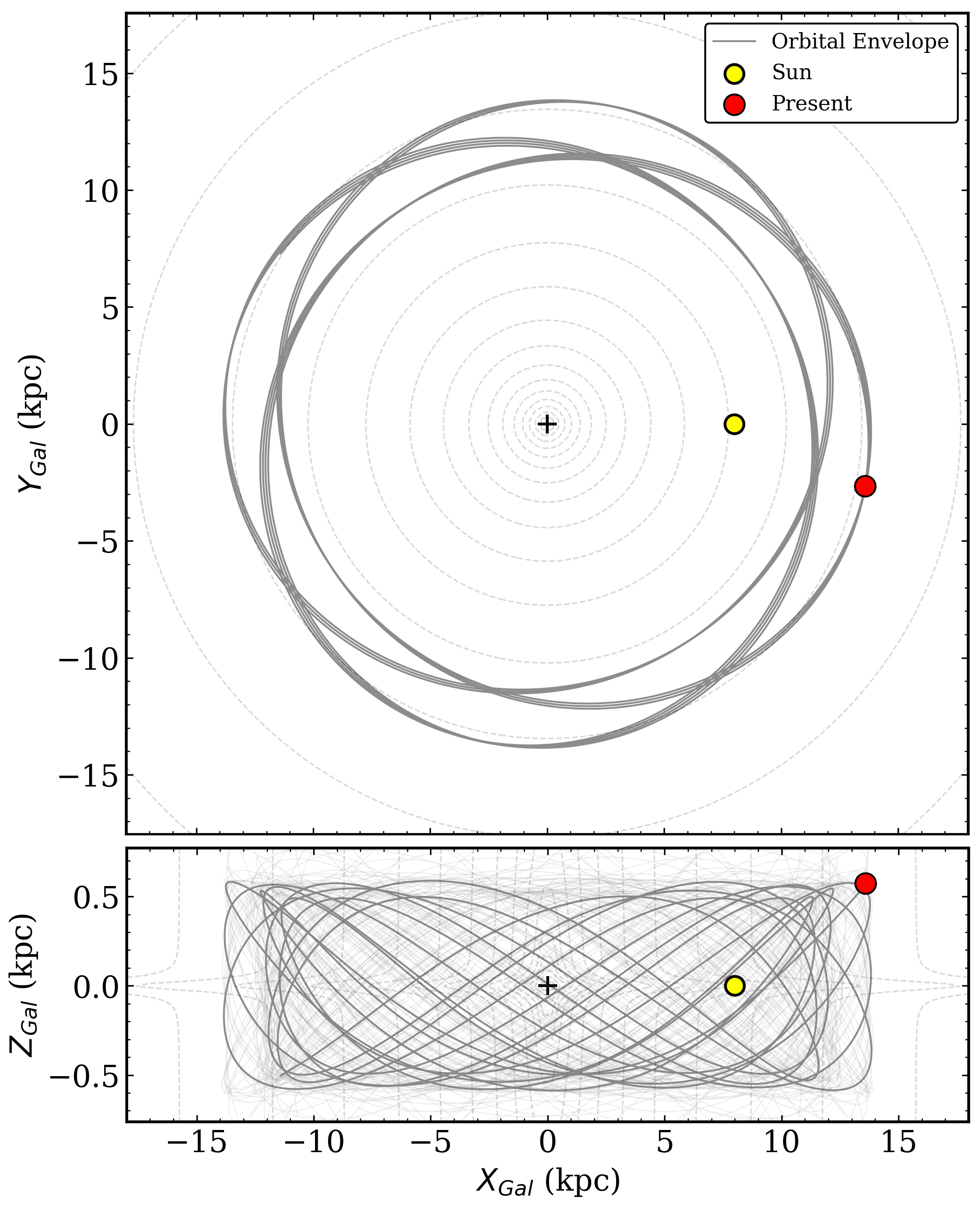}
   \caption{Reconstructed Galactic orbit of Berkeley~31. \textit{Top:} projection onto the $X_{\mathrm{Gal}}$--$Y_{\mathrm{Gal}}$ plane. \textit{Bottom:} projection onto the $X_{\mathrm{Gal}}$--$Z_{\mathrm{Gal}}$ plane. The thin gray curves represent the orbits obtained from the Monte Carlo realizations of the observational uncertainties. The present-day position of Berkeley~31 and the position of the Sun are marked by filled circles.}
\label{fig:orbit_b31}
\end{figure}

The orbit was calculated from the cluster's sky position, distance, proper motion, and systemic radial velocity. We used $R_0=8.2\pm0.1$~kpc for the Galactocentric distance of the Sun, a circular velocity of $V_{\rm rot}=220$~km~s$^{-1}$, and a solar vertical position of $Z_0=25\pm5$~pc \citep{Bovy2012}. The transformation to Galactocentric coordinates followed the prescription of \citet{Tuncel2019}. The cluster orbit was integrated both backward and forward in time with a timestep of 1~Myr over the adopted cluster age of 3.7~Gyr. The effects of the Galactic bar and spiral arms were also tested using the corresponding non-axisymmetric potentials available in \texttt{galpy}. These additional integrations did not produce significant changes in the main orbital parameters.

To account for the uncertainties in the observed phase-space parameters, we generated Monte Carlo realizations of the input quantities and integrated the orbit for each realization. From the resulting distributions, we derived the perigalactic distance ($R_{\rm peri}$), apogalactic distance ($R_{\rm apo}$), orbital eccentricity ($e$), maximum vertical distance ($Z_{\rm max}$), guiding radius ($R_{\rm g}$), and the radial, vertical, and azimuthal orbital periods. The derived orbital parameters are listed in Table~\ref{tab:summary}, and the reconstructed orbit is shown in Figure~\ref{fig:orbit_b31}.

The Galactic space-velocity components $(U, V, W)$ were calculated from the cluster's mean proper motion, distance, and systemic radial velocity using the \texttt{galpy} package. To correct for the influence of Galactic differential rotation on the in-plane velocity components, we adjusted the $U$ and $V$ components according to the formulation given by \citet{Mihalas1981}, while leaving the $W$ component uncorrected. The heliocentric velocity components were then corrected for the solar motion, adopting $(U, V, W)_{\odot}=(8.83\pm0.24,14.19\pm0.34,6.57\pm0.21)$~km~s$^{-1}$ \citep{Coskunoglu2011}. The total space velocity with respect to the local standard of rest was calculated as $S_{\rm LSR}=\sqrt{U_{\rm LSR}^{2}+V_{\rm LSR}^{2}+W_{\rm LSR}^{2}}$. The resulting space-velocity components and total LSR velocity, together with the orbital parameters, are listed in Table~\ref{tab:summary}.

\subsection{Mass Segregation}
\label{sec:mass_seg}

\begin{figure}
    \centering
    \includegraphics[width=0.95\linewidth]{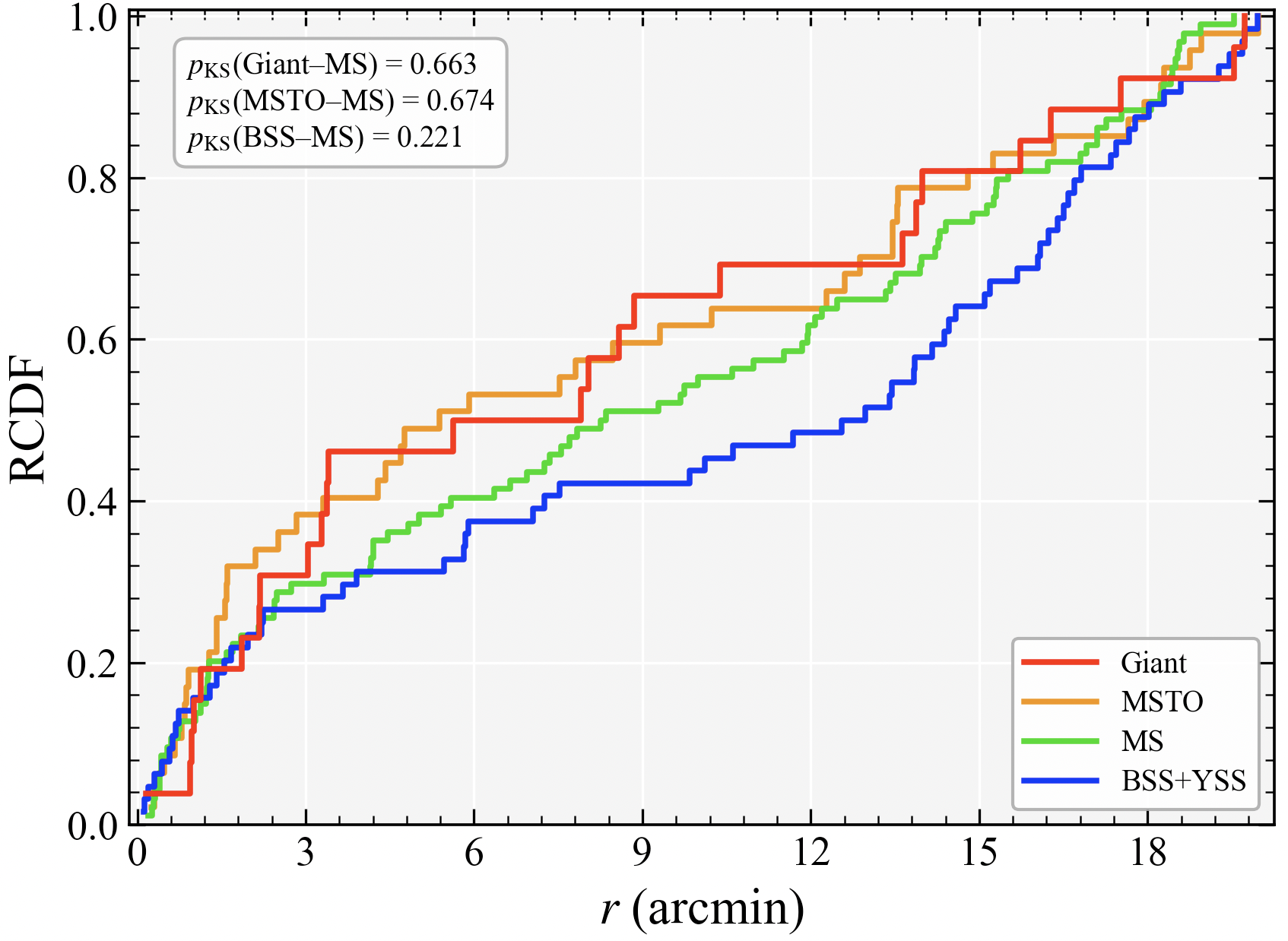}
    \caption{RCDFs for the giant, MSTO, MS, and BSS populations of Berkeley~31. The radial distance is the projected angular separation from the cluster center in arcminutes. We selected the populations based on their $G$-band magnitude ranges and membership probabilities. The KS-test $p$-values for selected population pairs are shown in the upper-left corner.}
    \label{fig:rcdf_berkeley31}
\end{figure}

We examined mass segregation by comparing the radial distributions of stars at different evolutionary stages. We divided the 329 probable members of Berkeley~31 into four mutually exclusive groups based on their positions in the CMD: 36 giant stars, 40 MSTO stars, 189 MS stars, and 64 BSS+YSS candidates. We defined the giant, MSTO, and MS populations by their $G$-band magnitude ranges, while treating BSS+YSS candidates separately. Figure~\ref {fig:rcdf_berkeley31} shows the radial cumulative distribution functions (RCDFs) of these populations.

The RCDFs reveal some differences in the spatial distributions of the stellar populations. The giant and MSTO populations show similar radial distributions and are relatively more concentrated toward the cluster center, whereas the MS population extends more gradually toward larger radii. This behavior is qualitatively consistent with mass segregation, in which more massive populations are preferentially concentrated toward the inner regions. In contrast, the BSS+YSS population does not show the expected central concentration of a dynamically segregated high-mass population. Its RCDF remains below that of the MS population over a substantial radial range and rises more strongly toward the outer regions, indicating a relatively extended distribution. However, none of the pairwise KS tests yield a statistically significant difference, with the smallest $p$-value obtained for the BSS+YSS--MSTO comparison ($p=0.0715$). Thus, while the RCDFs suggest a possible mass-dependent spatial structure, the present data do not provide statistically significant evidence for mass segregation in Berkeley~31.

\section{Stellar Populations and Evolutionary Features}
\label{sec:evolutionary_populations}

\subsection{Unresolved Binary Fraction}
\label{subsec:binary_fraction}

The binary fraction of an OC provides information on its dynamical evolution, internal structure, and long-term stability \citep[e.g.][]{vonHippel2002, Sollima2007}. Unresolved binaries can appear brighter than single stars of the same mass, broadening the MS or producing a secondary sequence in the CMD. Accounting for these systems is therefore important when characterizing an open cluster's stellar population \citep{Milone2012}. Following the photometric approach of \citet{Milone2012} and \citet{Donada2023}, we investigated the unresolved binary population of Berkeley~31.

\begin{figure*}
    \centering
    \includegraphics[width=1\linewidth]{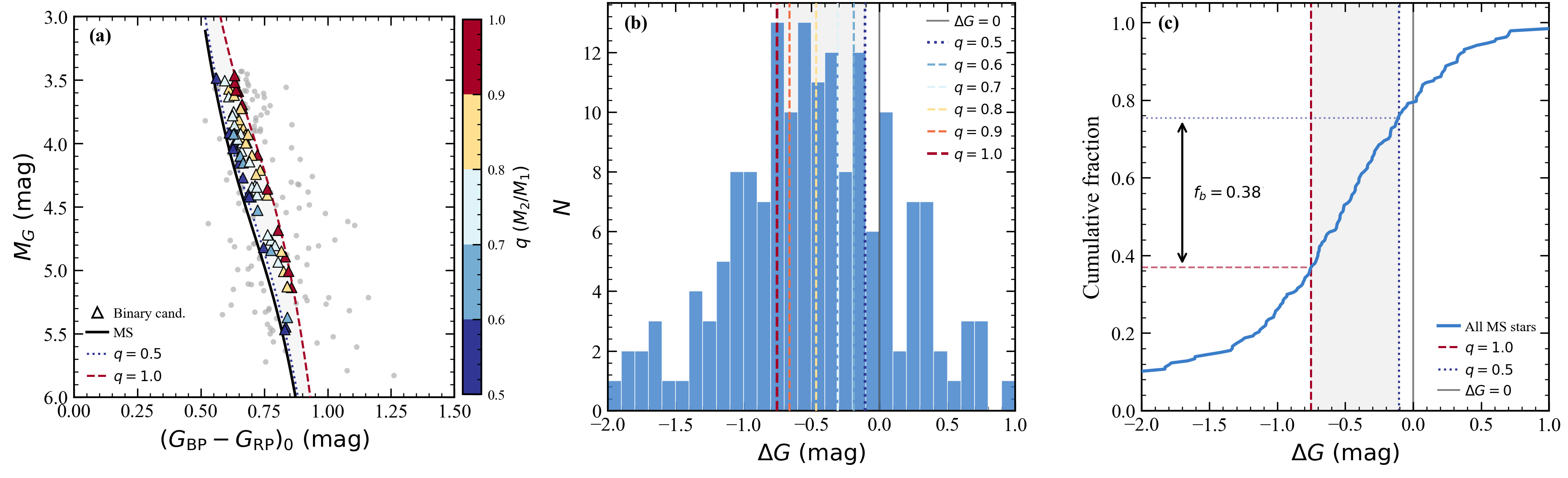}
     \caption{Binary fraction analysis of Berkeley~31. (a) De-reddened CMD of the selected MS stars. The solid line shows the fitted single-star ridge line, while the dotted and dashed lines indicate the binary-selection boundaries for $q=0.5$ and $q=1.0$, respectively. Triangle symbols mark the identified binary candidates, with their colors representing the corresponding mass ratios. (b) Distribution of the $\Delta G$ residuals relative to the single-star ridge line. The vertical lines indicate the expected magnitude offsets for different mass ratios. (c) Cumulative distribution of $\Delta G$ for the analyzed MS population.}
    \label{fig:berk31_binary}
\end{figure*}

MS stars were selected from the MSTO region down to four magnitudes below the MSTO, yielding 189 MS stars. We fitted a fourth-degree polynomial to their median ridge line to define the single-star sequence. The photometry was corrected for the adopted reddening and distance modulus listed in Table~\ref{tab:summary}, and the analysis was performed in the de-reddened CMD.

For an unresolved binary, the magnitude difference from a single star with the same primary mass is given by $\Delta m=-2.5\log_{10}(1+q^{\alpha})$, where $q=M_2/M_1$ is the mass ratio and $\alpha=3.5$ is the adopted mass--luminosity exponent for MS stars \citep{Eker2018}. For $q=0.5$ and $q=1$, the expected offsets are $\Delta m\simeq-0.092$ and $-0.753$ mag, respectively. We therefore classified stars between these two limits as binary candidates.

Figure~\ref{fig}a shows the de-reddened CMD, the fitted single-star ridge line, and the binary-selection boundaries. The binary candidates are shifted toward brighter magnitudes relative to the single-star sequence. Figure~\ref{fig:berk31_binary}b shows the $\Delta G$ residuals from the ridge line, with vertical lines indicating the expected offsets for different mass ratios. The corresponding cumulative distribution and the fractions within the $q=1$ and $q=0.5$ boundaries are shown in Figure~\ref{fig:berk31_binary}c.

Of the 189 MS stars, 72 lie within the binary-candidate region, giving $f_{\rm b}=0.38$. The binomial uncertainty is $\pm0.036$ at the $1\sigma$ level, resulting in $f_{\rm b}=0.38\pm0.04$. Thus, about 39\% of the analyzed MS population is consistent with unresolved binaries with $q\gtrsim0.5$. Since this photometric method becomes less sensitive to lower mass ratios, the measured fraction should be regarded as a lower limit to the intrinsic binary fraction of Berkeley~31.

The mass-ratio distribution is non-uniform, with a concentration toward relatively high $q$ values and a pronounced excess around $q\simeq0.8$--$0.9$. The lower-$q$ component may include different binary configurations, such as semi-detached or interacting systems \citep[e.g.][]{Bilir2005, Demircan2006, Ibanoglu2006, Eker2014}. Spectroscopic observations are needed to distinguish these populations and determine their orbital properties.

\subsection{Blue Straggler and Yellow Straggler Stars}
\label{subsec:bss_yss}

We identified 62 BSS and 2 YSS candidates among the probable members of Berkeley~31. \citet{Jadhav2021bss} previously reported 15 BSS candidates in Berkeley~31, 14 of which are recovered in our membership-selected sample. In addition, we identify 48 BSS candidates that were not reported in that study, together with two YSS candidates presented for the first time in this study. Thus, the present analysis substantially expands the previously identified BSS population of Berkeley~31. The BSS candidates in common with \citet{Jadhav2021bss} are shown in the CMD in Selection~\ref{sec:comp_membership}, while the spatial distribution of the BSS and YSS candidates within Berkeley~31 is shown in Figure~\ref{fig:lit_comparison}.

An important contribution of the present study is the extension of the previous ultraviolet investigation of Berkeley~31 BSSs. In \citet{Cinar2026b}, we investigated a subset of 11 BSSs with available \textit{Swift}/UVOT observations and searched for ultraviolet excesses associated with hot compact companions. These previously analyzed systems are retained here to provide continuity with the earlier study, while the present work adds 17 additional BSSs with \textit{Swift}/UVOT measurements. Consequently, the UV-based analysis now covers 28 BSS candidates in Berkeley~31.

\subsubsection{SED Analysis for Hot Component Searches}
\label{sec:sed}

We constructed spectral energy distributions (SEDs) for the 62 BSS and two YSS candidates identified in Berkeley~31. The photometric data used in the SED analysis cover the ultraviolet, optical, and infrared wavelength ranges, as described in Section~\ref{sec:data}. Among the 64 candidates, 28 have available \textit{Swift}/UVOT measurements. Of these, the SEDs of 11 stars had already been analyzed in our previous study \citet{Cinar2026b}. In the present work, we investigate the possible presence of hot secondary components in 17 BSSs for the first time. The UV measurements were used to identify excess emission relative to the single-component SEDs, while the optical and infrared photometry constrained the BSS primaries. The candidates showing significant UV excesses were then examined using two-component SED models that included the available ultraviolet measurements. The remaining BSS candidates without UV measurements were also included in the SED analysis using their available optical and infrared data.

For each BSS candidate, the multi-band photometric measurements were analyzed with the Virtual Observatory SED Analyzer (VOSA; \citealt{Bayo2008}). The cluster distance and extinction adopted in the SED analysis were taken from the fundamental parameters derived in Section~\ref{sec:distance_reddening}. VOSA was used to compare the observed photometric fluxes with theoretical stellar atmosphere models and to determine the best-fitting single-star model. The photometric data were corrected for interstellar extinction using the adopted cluster reddening and extinction law.

\begin{figure*}
    \centering
    \includegraphics[width=0.9\linewidth]{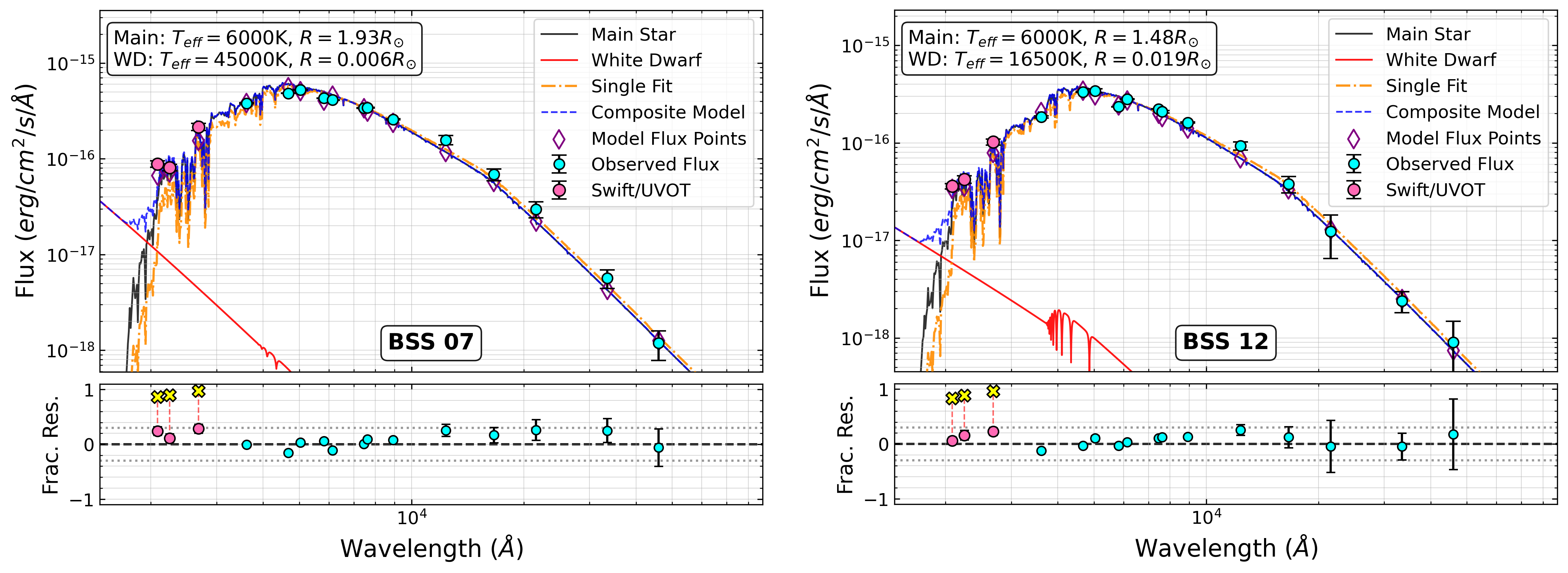}

    \caption{Two-component SED fits for the two binary BSS candidates. \textbf{Top panels:} Cyan circles show the observed photometric fluxes with their uncertainties. The black and red solid lines represent the best-fitting spectra of the cool BSS primary \citep{Castelli1997} and the hot WD companion \citep{Koester2010}, respectively. The blue dashed line gives the combined model obtained from the two components, while the orange line shows the single-component fit. Purple diamonds mark the model fluxes integrated over the corresponding photometric bands. \textbf{Bottom panels:} Residuals between the observed fluxes and the composite model. The yellow crosses show the \textit{Swift}/UVOT residuals with respect to the single-star model of the BSS primary, highlighting the excess emission in the ultraviolet.}
    \label{fig:WD_BSS}
\end{figure*}

\begin{table*}[ht]
\centering
\footnotesize
\caption{Best-fitting parameters from the two-component SED models of the binary BSS candidates in Berkeley~31. The parameters of the primary (A) and hot companion (B) are listed together for each system.}
\label{tab:sed_results}
\renewcommand{\arraystretch}{1}
\setlength{\tabcolsep}{3pt}
\begin{tabular}{cccccc ccc ccc}
\hline
BSS ID & $\alpha$ & $\delta$ &
$T_{\rm eff,A}$ & $R_{\rm A}$ & $L_{\rm A}$ &
$T_{\rm eff,B}$ & $R_{\rm B}$ & $L_{\rm B}$ &
$N_{\rm fit}$ & $\chi^2_\nu$ & vgf$_b$ \\
 & (hh:mm:ss)& (hh:mm:ss) &
(K) & ($R_\odot$) & ($L_\odot$) &
(K) & ($R_\odot$) & ($L_\odot$) &
 & & \\
\hline

07~ & 06:57:37.16 & +08:16:59.2 &
6000$\pm$341 & 1.93$\pm$0.32 & 4.68$\pm$1.61 &
45,000$\pm$2500 & 0.006$\pm$0.001 & 0.065$\pm$0.022 &
18 & 1.93 & 8.19 \\

12~ & 06:57:29.90 & +08:18:21.5 &
6000$\pm$246 & 1.48$\pm$0.25 & 2.78$\pm$0.96 &
16,500$\pm$125 & 0.019$\pm$0.003 & 0.025$\pm$0.009 &
18 & 6.03 & 5.95 \\

26$^{*}$ & 06:57:37.44 & +08:17:16.8 &
8000$\pm$121 & 1.45$\pm$0.09 & 7.71$\pm$0.79 &
15,500$\pm$125 & 0.20$\pm$0.01 & 2.08$\pm$0.21 &
20 & 4.35 & 0.65 \\

27$^{*}$ & 06:57:40.56 & +08:17:45.6 &
8000$\pm$150 & 1.60$\pm$0.10 & 9.47$\pm$0.97 &
15,250$\pm$125 & 0.12$\pm$0.01 & 0.72$\pm$0.07 &
20 & 3.09 & 1.18 \\

28$^{*}$ & 06:57:36.24 & +08:17:24.0 &
7000$\pm$177 & 1.37$\pm$0.09 & 4.06$\pm$0.42 &
15,250$\pm$125 & 0.07$\pm$0.01 & 0.22$\pm$0.02 &
18 & 1.74 & 2.51 \\

\hline
\end{tabular}

\begin{flushleft}
\scriptsize
$^{*}$ Results adopted from \citet{Cinar2026b}.
\end{flushleft}
\end{table*}

The single-component SEDs were fitted with the Kurucz ODFNEW/NOVER atmosphere models \citep{Castelli1997,Castelli2003}, covering the effective-temperature range of $5,000$--$15,000$~K. The best-fitting model was selected by minimizing the reduced chi-square, $\chi^2_\nu$, between the observed and model fluxes. We also considered the visual goodness-of-fit parameter provided by VOSA, $V_{\rm gfb}$, which accounts for possible underestimation of the photometric uncertainties by imposing a minimum relative flux uncertainty. Following the criterion adopted in previous studies \citep[e.g.,][]{Jimenez-Esteban2021,Rebassa-Mansergas2019,Solano2021,Zeng2025}, fits with $V_{\rm gfb}<10$--15 were considered acceptable and retained for further analysis. \{According to the VOSA documentation\footnote{\url{https://svo2.cab.inta-csic.es/theory/vosa/help/star/intro/}}, $V_{\rm gfb}$ is a modified reduced chi-square statistic calculated by imposing a minimum relative uncertainty of 10\% on the observed fluxes. This reduces the influence of photometric measurements with potentially underestimated uncertainties. Values of $V_{\rm gfb}$ below approximately 10--15 are commonly considered indicative of an acceptable fit; however, this criterion does not replace the conventional reduced chi-square or guarantee that all model residuals are small.

For the 28 BSS candidates with \textit{Swift}/UVOT measurements, we examined the residuals between the observed UV fluxes and the best-fitting single-star models. The fractional residual was calculated as $(F_{\rm obs}-F_{\rm model})/F_{\rm obs}$, with particular attention given to the $UVM2$ and $UVW2$ bands. The UV excess of approximately 30\% or more in at least two \textit{Swift}/UVOT filters was considered evidence for a possible additional hot component \citep{Chand2024}. This criterion was used to separate significant UV excesses from deviations that may result from photometric uncertainties \citep{Sheikh2024, 2024AJ....168..274S, Jadhav2019blue, Bisht2026b}.
 
The UV excess criterion identified two additional BSS candidates with excess emission in at least two \textit{Swift}/UVOT bands. We fitted these two systems with two-component SED models, using Kurucz atmospheres for the BSS primaries and Koester WD models \citep{Castelli1997, Koester2010} for the hot components. The resulting SED fits are shown in Figure~\ref{fig:WD_BSS}. The fitted parameters indicate that the hot component of BSS~07 has a temperature of about 45,000~K, while that of BSS~12 is about 16,500~K. Their small radii and high temperatures indicate compact hot companions. Based on their high effective temperatures and very small inferred radii, the hot components are consistent with compact degenerate companions. The companion to BSS 07, with $T_{\rm eff}=45,000\pm2,500$ K and $R=0.006\pm0.001\,R_{\odot}$, is consistent with a hot white dwarf (WD), whereas the cooler component in BSS 12, with $T_{\rm eff}=16,500\pm125$ K and $R=0.019\pm0.003\,R_{\odot}$, is consistent with a pre-extremely-low-mass (pre-ELM) WD. We therefore regard these objects as hot WD and pre-ELM WD candidates rather than confirmed evolutionary classifications. Similar UV-based SED analyzes have been used to identify low-mass and extremely low-mass WD companions to BSSs in OCs \citep[e.g.,][]{Jadhav2019blue, 2021MNRAS.507.2373P}.

The two-component fit to BSS~12 has a higher reduced chi-square ($\chi_\nu^2=6.03$) than BSS~07 ($\chi_\nu^2=1.93$), although its $V_{\rm gfb}$ value of 5.95 is within the commonly accepted range. The high $\chi_\nu^2$ may reflect underestimated photometric errors, survey calibration differences, uncertainties in extinction, or limitations of the atmosphere models. The available data do not allow us to identify the main cause. Since the uncertainties reported by VOSA may not account for these systematic effects, the fitted temperature and radius of the BSS~12 companion should be treated with caution. Although the results are consistent with a hot compact object, its classification as a pre-ELM WD remains tentative pending independent confirmation.
 
Together with the 11 systems analyzed in \citet{Cinar2026b}, the present analysis brings the total number of BSSs with evidence for hot companions in Berkeley~31 to five. These systems include one hot WD and four pre-ELM WD candidates. The detection of such hot, compact companions is consistent with the presence of post-mass-transfer systems among BSSs, as found in other OCs through ultraviolet observations and SED modeling \citep[e.g.,][]{Jadhav2019blue,2021MNRAS.507.2373P}. The present results, therefore, add two new hot-component candidates to the Berkeley~31 BSS population.

For the BSS candidates without \textit{Swift}/UVOT measurements, we also checked the available observations from GALEX, eROSITA, Chandra, and XMM-Newton. We found no useful observations for these stars in these datasets. Therefore, the absence of a detected UV or X-ray excess does not provide sufficient evidence to classify these systems as single stars. In other words, the lack of a detected secondary component cannot be taken as evidence that no companion is present, particularly when suitable ultraviolet observations are unavailable. We therefore do not use the single-component SED fits to make a statement about the binarity of these stars. For completeness, we provide the single-component SED results for these BSS candidates in Appendix~\ref{app:single_sed}.
Identifying a hot secondary component from broadband SEDs is subject to several systematic uncertainties, including the adopted extinction, photometric calibration, possible source blending within the UVOT point-spread function, and limitations of the stellar-atmosphere models. We therefore used the UV excess criterion only as a selection step and did not interpret it independently as proof of a degenerate companion. Instead, we subsequently examined the candidate systems using two-component SED models. The inferred companions are considered plausible hot compact objects because the excess emission is detected in multiple ultraviolet bands and the resulting fits require components with high effective temperatures and very small radii. Nevertheless, spectroscopic or time-domain confirmation will be required to establish their evolutionary nature and binary orbital parameters.

\subsection{Variable-star Content}
\label{subsec:variables}

\begin{figure*}[!ht]
    \centering
    \includegraphics[width=0.72\linewidth]{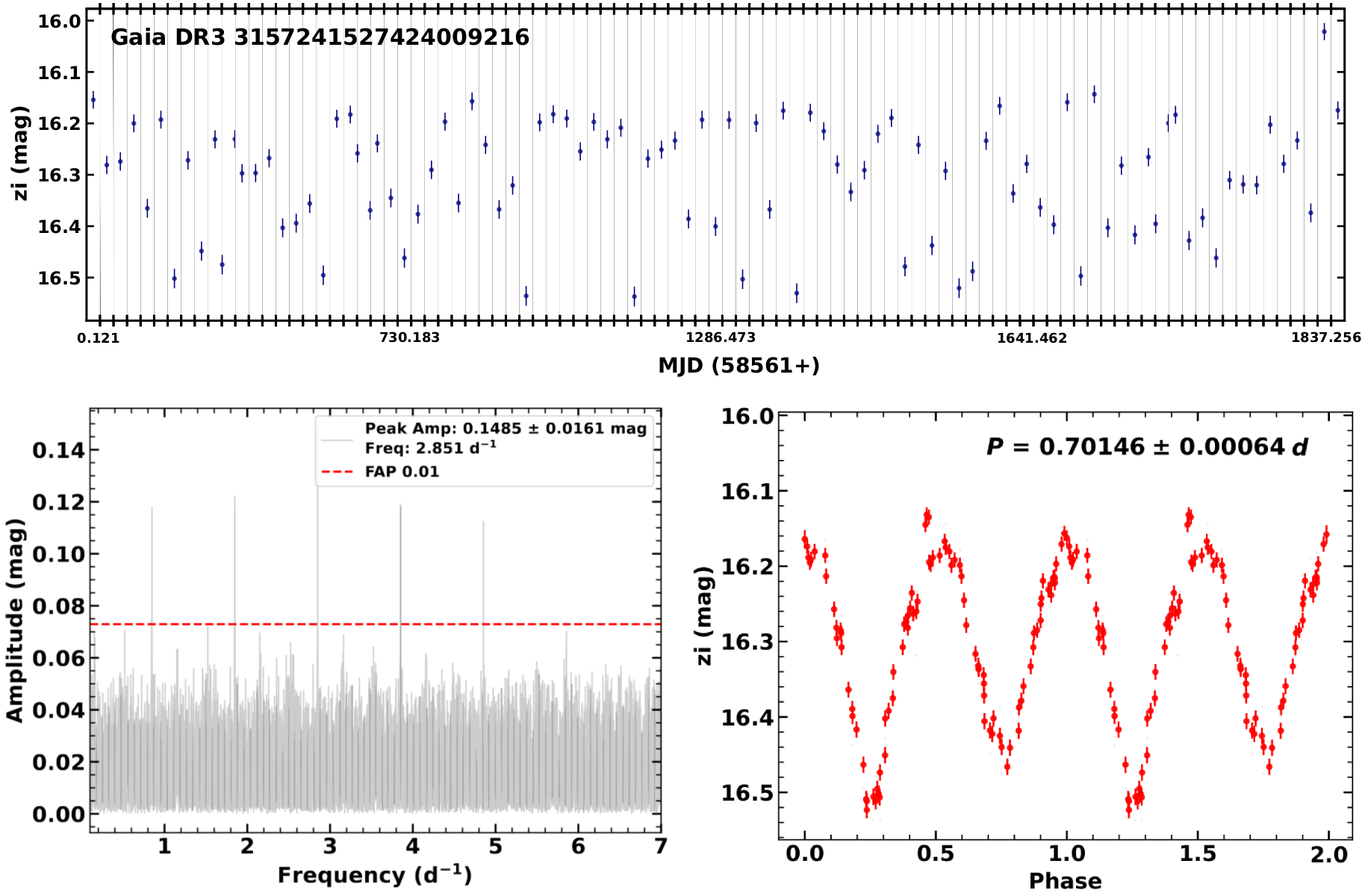}
    \caption{Time-series photometric analysis of the eclipsing binary candidate Gaia DR3 3157241527424009216. The top panel shows the observed $z$-band light curve versus MJD. The bottom-left panel presents the Lomb--Scargle periodogram, with the dominant frequency marked by the red dashed line and the adopted false-alarm probability (FAP) level indicated. The bottom-right panel shows the light curve phase-folded with the derived orbital period of $P = 0.70146 \pm 0.00064$ d. Two distinct minima per orbital cycle indicate an EW eclipsing binary.}
    \label{fig:lc_phase}
\end{figure*}

We searched for known variable stars among the Berkeley~31 members with $P\geq0.6$ by cross-matching our membership catalog with the Gaia~DR3 variability catalog \citep[VizieR I/358/vclassre;][]{Gaia_Variability}. Using a matching radius of $1^{\prime\prime}$, we identified two variable stars in common with the Gaia~DR3 catalog. Gaia DR3 source \texttt{3157241527424009216} ($G$=16.35 mag; $G_{\rm BP}$-$G_{\rm RP}$=0.59 mag) is classified as an eclipsing binary (ECL), while source \texttt{3157249155285881472} ($G$=14.99 mag; $G_{\rm BP}$-$G_{\rm RP}$=0.63 mag) is classified as a combination of $\delta$~Sct, $\gamma$~Dor, and SX~Phe-type variability. The two identified variables provide additional targets for studies of binary and pulsating-star properties in Berkeley~31. Given the small number of matches, we draw no further conclusions about the cluster's variable-star population here.

In this study, we used time-series photometric data for Gaia DR3 \texttt{3157241527424009216}, obtained from the Zwicky Transient Facility in the $zi$ band. The available data for the other source, Gaia DR3 \texttt{3157249155285881472}, were insufficient to reliably assess its variability, so we excluded it from the further analysis. For Gaia DR3 source \texttt{3157241527424009216} we obtained 119 observations and selected 95 of good quality. Time-series photometry of variable stars provides important information about stellar variability and can be used to investigate both intrinsic and extrinsic variations. Intrinsic variability relates to physical processes within the star, while extrinsic variability generally arises from geometric effects, such as eclipses in binary systems. For binary stars, light-curve analysis can provide fundamental parameters such as the orbital period and, with appropriate modeling, the stellar masses and radii.

Because BSSs can exhibit photometric variability, we examined available ZTF time-series data to investigate the BSS candidate's variability. The resulting light curve is shown in the upper panel of Figure \ref{fig:lc_phase}. To determine the binary system's periodicity, we performed a Lomb--Scargle periodogram analysis. The bottom-left panel of Figure \ref{fig:lc_phase} shows the periodogram (frequency versus amplitude). We selected significant frequencies using the criterion that the corresponding signal has an amplitude or power at least four times higher than the noise level, as suggested by \citet{breger1993nonradial}. We then used the dominant frequency from the periodogram to fold the light curve over the corresponding orbital period.

The resulting phase-folded light curve, shown in the bottom-right panel of Figure \ref{fig:lc_phase}, shows two distinct minima, characteristic of two eclipses in one orbital cycle. The derived orbital period is (0.70146 $\pm$ 0.00064) d, which is consistent with the period reported by \citet{chen2020zwicky}. The presence of two eclipses and the overall morphology of the phase-folded light curve are consistent with a W UMa-type eclipsing binary (EW type). This classification also matches the EW-type classification reported by \citet{chen2020zwicky}.

\section{Chemo-dynamical Evolution}
\label{subsec:birth_radius_migration}

We estimated the chemical birth radius ($R_{\rm b}$) of Berkeley~31 using the chemo-dynamical prescription of \citet{Minchev2018}, which accounts for the time evolution of the Galactic radial metallicity gradient. The cluster age and metallicity adopted for this calculation are listed in Table~\ref{tab:summary}. Since $R_{\rm b}$ depends on the adopted chemical-evolution model, it represents an approximate estimate of the cluster's birth location.

To characterize the present-day orbital position of Berkeley~31, we calculated its Galactocentric radius ($R_{\rm GC}$) and guiding radius ($R_{\rm g}$) from the derived kinematic parameters. We then used the differences $|R_{\rm g}-R_{\rm b}|$ and $|R_{\rm GC}-R_{\rm g}|$ to quantify the displacement between the inferred birth location, guiding center, and present-day Galactocentric position. The first quantity measures the change in guiding radius relative to the inferred birth radius, while the second describes the radial excursion around the guiding center. The total displacement from the birth radius is given by $|R_{\rm GC}-R_{\rm b}|$. For Berkeley~31, we obtain $|R_{\rm g}-R_{\rm b}|=1.14$~kpc and $|R_{\rm GC}-R_{\rm b}|=2.70$~kpc. The corresponding difference between the present Galactocentric and guiding radii is therefore $|R_{\rm GC}-R_{\rm g}|=1.56$~kpc. These values indicate that the cluster's guiding center is displaced from its inferred birth radius, while its present Galactocentric position is also offset from the guiding center.

For a more homogeneous comparison, we consider Berkeley~32 and Berkeley~36, which were previously analyzed by our group using the same methodology \citep{Cinar2026d,Cinar2026c}. These OCs have $|R_{\rm g}-R_{\rm b}|=1.00$ and $5.04$~kpc, respectively, and $|R_{\rm GC}-R_{\rm b}|=1.29$ and $4.71$~kpc. This common methodological framework allows their radial-displacement properties to be compared more directly with those of Berkeley~31. The three OCs, therefore, occupy different regions of the age--radial displacement plane, as shown in Figure~\ref{fig}.

\begin{figure*}
    \centering
    \includegraphics[width=0.75\linewidth]{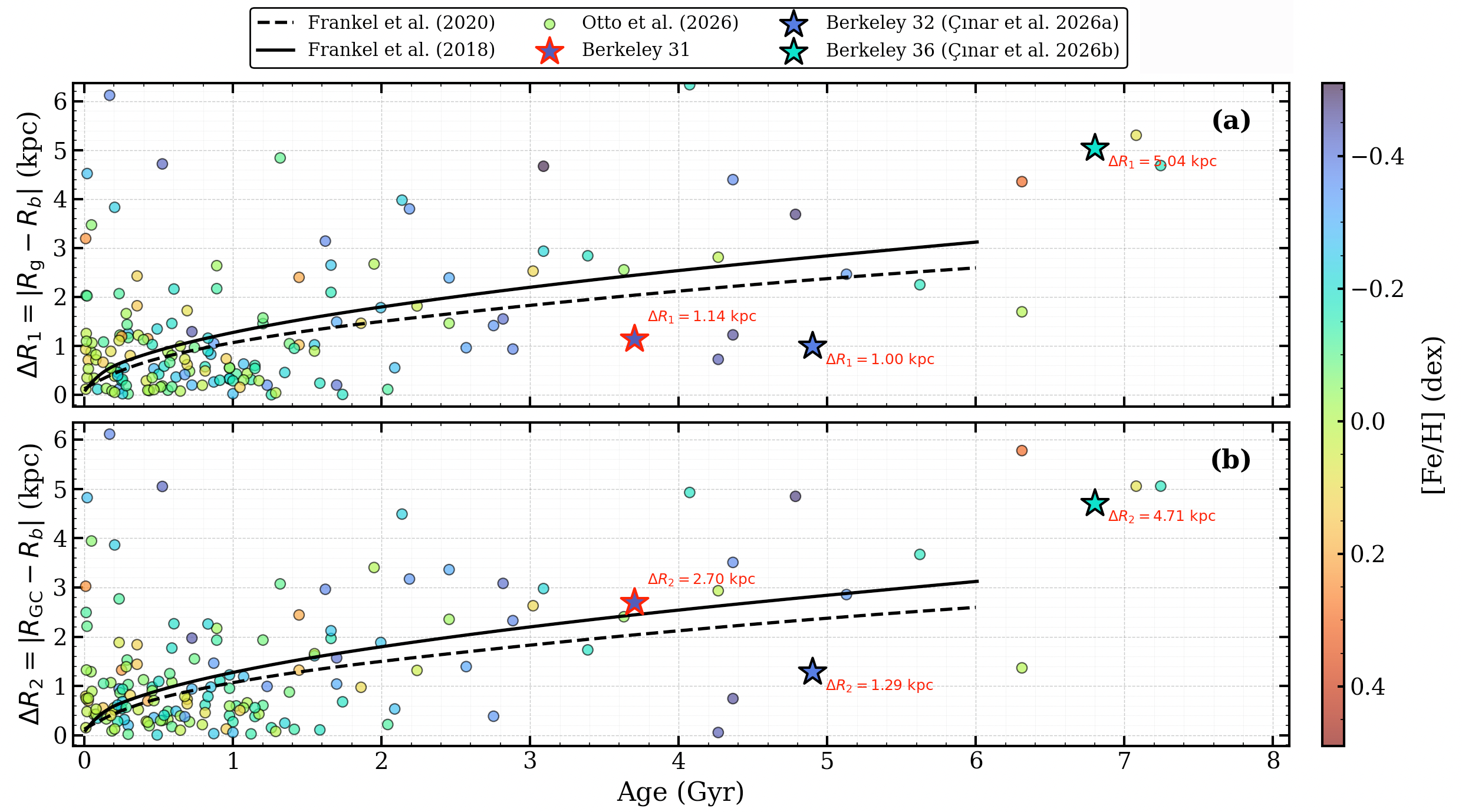}
    \caption{Distribution of OCs in the age--radial displacement plane. Panel (a) shows the absolute difference between the guiding radius and the birth radius, $|R_{\rm g}-R_{\rm b}|$, while panel (b) shows the absolute difference between the present Galactocentric radius and the birth radius, $|R_{\rm GC}-R_{\rm b}|$. The color scale represents the cluster metallicity, [Fe/H]. The background circles show the OCs from \citet{Otto2026}, while Berkeley~31, Berkeley~32, and Berkeley~36 are marked with star symbols. The values for Berkeley~32 and Berkeley~36 are taken from our previous studies \citep{Cinar2026d,Cinar2026c}. The solid and dashed curves indicate the empirical age-dependent relations from \citet{Frankel_2018,Frankel2020}.}
    \label{fig:rguide_rgc}
\end{figure*}

\section{Discussion}

The results presented in the previous sections provide a combined view of the stellar population, chemical properties, kinematics, and Galactic location of Berkeley~31. In this section, we place these results in the context of previous studies and discuss the evolutionary history of the cluster. We first compare our derived fundamental parameters with published values and then consider the chemical and dynamical properties of Berkeley~31 in relation to its present Galactocentric position and possible radial migration.

\subsection{Comparison with Previous Studies}
\label{subsec:literature_context}

\begin{figure}
    \centering
    \includegraphics[width=0.9\linewidth]{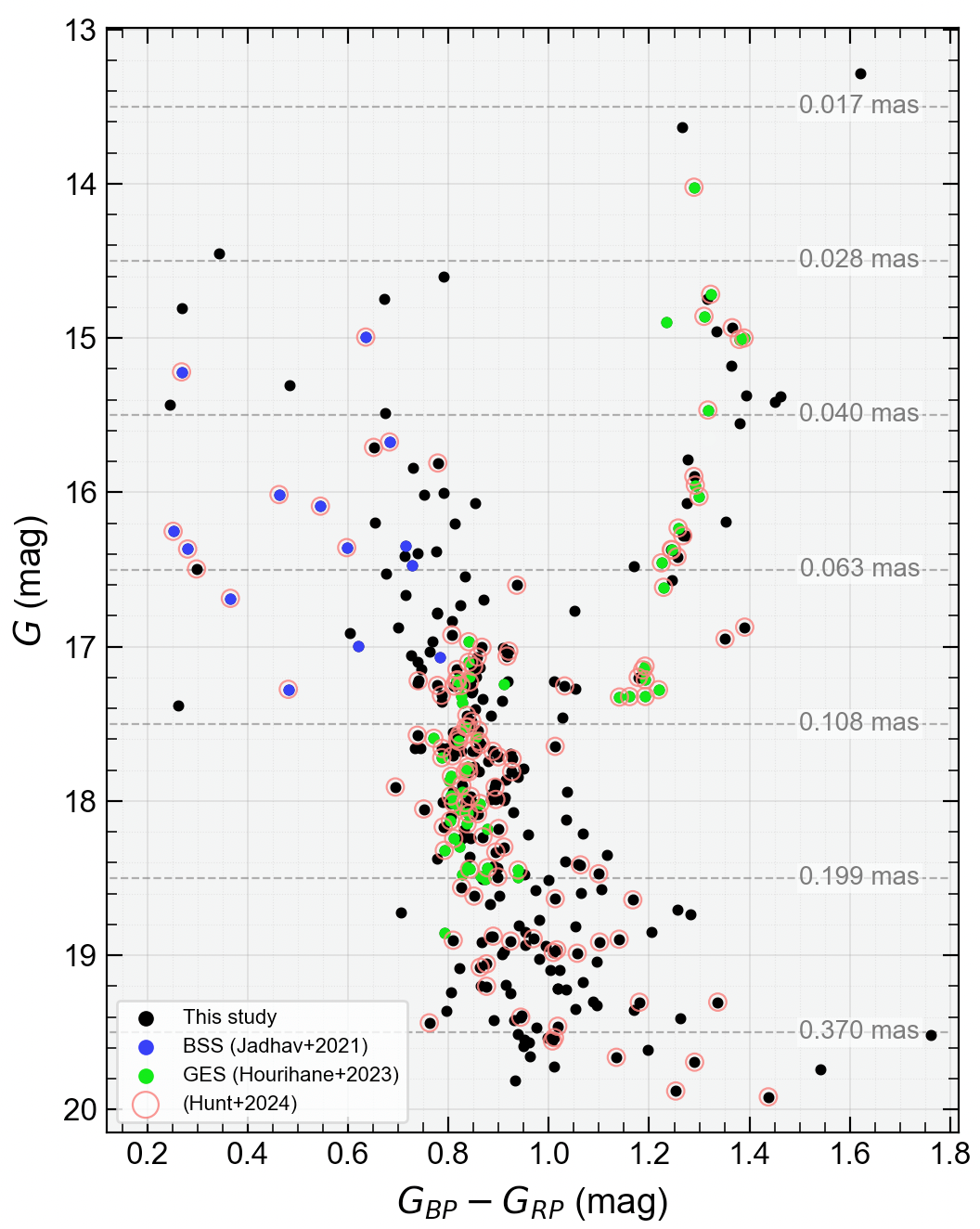}
    \caption{CMD of Berkeley~31 used to compare the membership sample adopted in this study with previous catalogs.
    Black points show the members identified from this study, while blue and green points indicate the BSS candidates reported by\citet{Jadhav2021bss} and \citet{Hourihane2023}, respectively. Open red circles mark the stars listed as members by \citet{Hunt2023}. The gray dashed lines indicate the mean parallax uncertainty calculated in 1-mag $G$-magnitude intervals.}
    \label{fig:B31_CMD_membership}
\end{figure}
\noindent$\bullet$\quad\textbf{Cluster Membership:}
\label{sec:comp_membership} 
The Gaia CMD of Berkeley~31 is shown in Figure~\ref{fig:B31_CMD_membership}. The stars selected as members on the basis of their Gaia astrometric properties are represented by black points. The membership list of \citet{Hunt2023} was not adopted directly; instead, membership was redetermined using the same astrometric selection procedure applied to the other clusters in this study. In this way, a homogeneous member sample was constructed using the same Gaia quantities and selection criteria. The members identified by \citet{Hunt2023} are shown by open red circles for comparison. Previously identified BSS candidates are also indicated: those reported by \citet{Jadhav2021bss} are marked in blue, while the candidates listed in the Gaia-ESO Survey catalog of \citet{Hourihane2023} are shown in green. The gray dashed lines in Figure~\ref{fig:B31_CMD_membership} show the mean trigonometric parallax uncertainties calculated in 1-mag bins over $13\leq G {\rm~(mag)}<20$. The larger uncertainties at fainter magnitudes limit the ability of parallax measurements to distinguish cluster members from field stars and should therefore be considered when interpreting the membership selection.
\noindent

$\bullet$\quad\textbf{Astrometric Parameters:}
\label{sec:astrometric_comparison} 
The adopted center for Berkeley~31, $(\alpha,\delta)=(06^{\rm h}57^{\rm m}37^{\rm s}.45,+08^{\rm o}17^{'}13^{''}.70)$, is found to be close to the positions listed by \citet{Cantat-Gaudin20}, $(104^\circ.406,8^\circ.285)$, and \citet{Hunt2023}, $(104^\circ.407,8^\circ.288)$. Similarly, the mean proper motion, $(\langle\mu_{\alpha}\cos\delta\rangle,\langle\mu_{\delta}\rangle)=(0.110\pm0.215,-0.920\pm0.186)$~mas~yr$^{-1}$, is found to be close to the values reported by \citet{Cantat-Gaudin20}, $(0.169,-0.926)$~mas~yr$^{-1}$, and \citet{Hunt2023}, $(0.114,-0.922)$~mas~yr$^{-1}$. Thus, good agreement is found for both the cluster position and proper motion.

The mean trigonometric parallax, $\langle\varpi\rangle=0.166\pm0.005$~mas, is found to be higher than the values of 0.141 and 0.123~mas reported by \citet{Cantat-Gaudin20} and \citet{Hunt2023}, respectively. The corresponding differences are estimated to be 0.025 and 0.043~mas. Their statistical significance cannot be assessed directly because uncertainties for the literature values are not provided in the Unified Cluster Catalog \citep[UCC,][]{Perren2023}. These differences may also be partly attributed to differences in member selection and parallax treatment. A total of 329 members are included in the present sample, whereas 375 stars are assigned membership probabilities above 50\% in the UCC. However, these numbers cannot be directly compared without matching the individual stars.

\begin{figure*}
    \centering
    \includegraphics[width=0.95\linewidth]{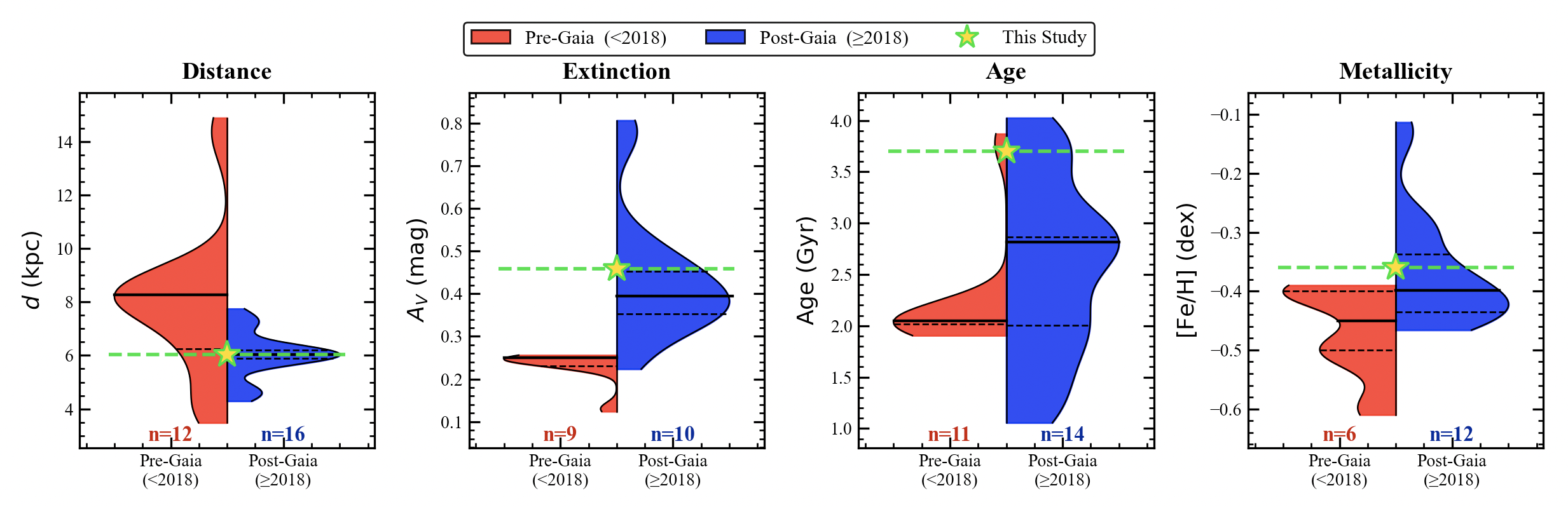}
    \caption{Literature estimates of the distance, $V$-band extinction, age, and iron abundance of Berkeley~31. The red and blue halves of each distribution show values from the pre-Gaia and post-Gaia studies, respectively. Black solid lines mark the medians, and black dashed lines mark the lower and upper quartiles. Yellow stars indicate the values derived in this study, with horizontal green dashed lines extending across each panel for comparison. The sample size of each literature group is given at the bottom of the panel.}
    \label{fig:lit_comparison}
\end{figure*}

\noindent$\bullet$\quad\textbf{Fundamental Parameters:}
\label{sec:fundamental_comparison} Figure~\ref{fig:lit_comparison} compares our estimates of the distance, extinction, age, and iron abundance of Berkeley~31 with values reported in the literature. We divide the published estimates into studies before 2018 (pre-Gaia) and those from 2018 onward (post-Gaia). This division is based on publication year and is used only to illustrate the distribution of published values. Since some studies may use the same catalogs and different analysis methods, the distributions should not be regarded as independent measurements.

A wider spread is seen in the distance estimates from the earlier studies, whereas the more recent values are found to be more concentrated around 6~kpc. The parallax-based distance, $d_{\varpi}=6.02\pm0.18$~kpc, is found to lie close to the central values of the later sample. A similar level of agreement is seen in the extinction estimates, with $\langle A_V\rangle=0.459\pm0.145$~mag being close to the higher values reported in recent studies.

A larger spread is seen in the age estimates than in the other parameters. The isochrone age of $3.7\pm0.2$~Gyr is located toward the older end of the published distribution, but is found to be consistent with the values reported by \citet{Dias2021} and \citet{Cavallo2024}. An independent age estimate of $3.72\pm1.42$~Gyr is obtained from the chemical clock and is also found to be consistent with the isochrone result.

A relatively broad range is covered by the published iron abundances, although a subsolar metallicity is indicated by most recent estimates. The value obtained in the present study, $[\mathrm{Fe/H}]=-0.36\pm0.04$~dex, is found to be consistent with this range and with the values reported by \citet{Fu2022}, \citet{Spina2021}, and \citet{Zhong2020}. The spread among the literature values further indicates that the difference between the earlier and more recent estimates should not be interpreted as a systematic change in the cluster metallicity.

Berkeley~31 is therefore placed at a distance of about 6~kpc, with moderate extinction, subsolar metallicity, and an age of about 3.7~Gyr. Among these parameters, the largest difference from the central values of the recent literature sample is found for the age, whereas the distance, extinction, and metallicity are found to be broadly consistent with previous studies.

\noindent$\bullet$\quad\textbf{Radial Velocity:}
\label{sec:Radial-velocity_comparison} 
The RV measurements reported for Berkeley~31 range from 42.8 to 65.3~km~s$^{-1}$, with most values lying between about 59 and 65~km~s$^{-1}$. A mean cluster radial velocity of $\langle V_{\rm rad,OC}\rangle=57.89\pm0.08$~km~s$^{-1}$ is derived in the present study. This value is close to the 58.8~km~s$^{-1}$ reported by \citet{Otto2026} and is broadly consistent with the measurements around 60~km~s$^{-1}$ reported by \citet{Donor2020} and \citet{Netopil2022}. Lower values of 55.2 and 42.8~km~s$^{-1}$ were reported by \citet{Fu2022} and \citet{Zhong2020}, respectively.

These differences are unlikely to arise primarily from the Gaia data release itself, since both \citet{Zhong2020} and \citet{Fu2022} adopted Gaia DR2-based cluster memberships, whereas the stellar RVs were obtained from low-resolution LAMOST spectroscopy using DR5 and DR8, respectively. Systematic offsets of about 5~km~s$^{-1}$ between LAMOST RVs and measurements from Gaia and higher-resolution spectroscopic surveys have been reported in several studies \citep{Anguiano2018, Zhong2020, Fu2022}. In particular, \citet{Zhong2020} found mean offsets of $-5.1$ and $-5.5$~km~s$^{-1}$ between their LAMOST-based cluster RVs and those from Gaia and APOGEE, respectively, while \citet{Fu2022} obtained a median difference of $V_{\rm rad, Gaia~DR3}-V_{\rm rad, LAMOST}=4.66$~km~s$^{-1}$. The known LAMOST RV offset is therefore comparable to, and could plausibly account for, the relatively small difference between the value derived here and the 55.2~km~s$^{-1}$ reported by \citet{Fu2022}. In contrast, the substantially lower value of 42.8~km~s$^{-1}$ reported by \citet{Zhong2020} cannot be explained by the LAMOST zero-point offset alone. The remaining discrepancy may instead reflect differences in the adopted member sample, the number of stars used to determine the cluster mean RV, or possible field-star contamination. For individual members, unresolved binarity may introduce additional RV shifts through orbital motion.

\subsection{Blue Straggler Stars and Binary Evolution}

The present study significantly enlarges the previously investigated BSS population of Berkeley~31. We identify 62 BSS candidates, of which 14 overlap with the BSS sample reported by \citet{Jadhav2021bss}, while 48 are newly identified in the present membership-selected sample. In addition, the ultraviolet analysis extends our previous study of 11 BSSs \citep{Cinar2026b}, incorporating 17 additional BSSs with available \textit{Swift}/UVOT observations. Among the 28 UV-observed BSSs considered in the combined analysis, five show evidence for hot compact companions. This expanded sample, therefore, provides a broader observational basis for investigating the frequency and nature of post-mass-transfer systems among the BSS population of Berkeley~31.

The Galactic orbital history of Berkeley~31 provides a broader context for the cluster, but it should not be interpreted as causally linked to the formation of its BSS population. Radial migration traces the long-term motion of the cluster within the Galactic disk, whereas the evolutionary origin of BSSs is governed by internal stellar and binary processes \citep{Echeveste2026b,Echeveste2026a}. In this respect, the detection of UV excesses associated with hot compact companions provides independent evidence for binary evolution among some BSS systems. Such excesses, when consistent with hot WD or pre-WD companions, support a past mass-transfer origin for at least a fraction of the BSS population. The orbital and UV analyzes, therefore, probe distinct physical processes and timescales: the former traces the long-term Galactic evolution of Berkeley~31, while the latter constrains the evolutionary histories of individual BSS systems.

\subsection{Radial Migration and the Dynamical History}
\label{subsec:discussion_migration}

The chemo-dynamical analysis places Berkeley~31 in an intermediate regime of radial evolution. Its guiding radius is displaced from the inferred chemical birth radius, while its present Galactocentric radius is also offset from the guiding center. The former is commonly associated with churning, which changes the guiding-center radius, whereas the latter reflects the radial excursions produced by epicyclic motion, or blurring \citep{Sellwood2002, Schoenrich2009}. The evolution of guiding radii can also be driven by dynamical perturbations, with tidal interactions providing one mechanism for radial migration \citep{Hebrail2025}. For Berkeley~31, both effects therefore contribute to its present radial displacement. The relatively similar values of $|R_{\rm g}-R_{\rm b}|$ for Berkeley~31 and Berkeley~32, together with their different $|R_{\rm GC}-R_{\rm g}|$ values, further illustrate that a similar displacement from the inferred birth radius does not necessarily arise from the same orbital process.

The comparison with Berkeley~32 and Berkeley~36 provides a broader context for the dynamical evolution of Berkeley~31. Berkeley~32 shows a relatively small change in guiding radius but a more noticeable offset between its present Galactocentric and guiding radii, indicating a stronger contribution from blurring. Berkeley~36, in contrast, has a much larger displacement between its birth and guiding radii, while its present Galactocentric radius remains relatively close to its guiding radius. Its large radial offset is therefore more closely associated with a change in guiding radius than with its present orbital phase \citep{Cinar2026d, Cinar2026c}. Berkeley~31 lies between these two cases, with both a moderate guiding-radius displacement and a substantial present-day radial excursion. The importance of distinguishing between different radial measures is also highlighted by \citet{Yontan2026}, who examined the radial metallicity gradient of King OCs using complementary Galactic radius definitions and dynamical constraints to reduce the effects of radial migration and orbital blurring. A complementary comparison is provided by \citet{Canbay2026}, who reported modest radial offsets for eight UBC OCs by comparing orbital traceback radii with present-day guiding radii, although their traceback estimates differ from the chemically inferred birth radii adopted here.

The positions of the three OCs in the age--radial displacement plane also provide a useful comparison with the empirical relations of \citet{Frankel_2018, Frankel2020}. Berkeley~31 and Berkeley~32 fall within the range covered by these relations, whereas Berkeley~36 lies above them. This difference suggests that the three clusters do not share the same degree or character of radial displacement. However, the inferred birth radius is model-dependent because it is derived from the cluster age and metallicity within a specific chemo-dynamical prescription. The radial offsets should therefore be interpreted as estimates of the clusters' dynamical histories rather than as direct measurements of their formation radii or migration paths.

The inferred radial displacement is subject to uncertainties and model dependence in the chemical birth-radius estimate. For Berkeley 31, the difference between the present guiding radius and the inferred birth radius is $R_{\rm g}-R_{\rm b}=1.14\pm0.58$ kpc, corresponding to a moderate, approximately $2\sigma$ offset. We therefore interpret the result as suggestive evidence for an outward change in the guiding radius rather than as a precise measurement of the migration distance. The larger difference between the present Galactocentric radius and the birth radius additionally includes the contribution from the cluster's current orbital phase. Consequently, the present analysis supports a scenario in which both guiding-center evolution and orbital blurring contribute to the observed radial displacement, while the exact migration history cannot be uniquely reconstructed from the present-day phase-space information alone.

\section{Conclusions}
\label{sec:conclusions}

We have presented a detailed multi-wavelength study of the open cluster Berkeley~31, combining Gaia~DR3 astrometry, Gaia-ESO spectroscopy, and ultraviolet-to-infrared photometry. We used the cluster members to characterize its fundamental properties, investigate its stellar populations, and examine its Galactic orbital evolution. Particular attention was given to the BSS population and the possible presence of hot compact companions through ultraviolet excesses and SED modeling.

\begin{itemize}

\item We identified 329 probable members of Berkeley~31 from their Gaia~DR3 astrometric properties. The cluster is located at a distance of $6024\pm181$~pc, with a reddening of $E(B-V)=0.148\pm0.047$~mag. The mean metallicity is [Fe/H] $=-0.36\pm0.04$~dex, while the mean $\alpha$-element abundance is $\langle[\alpha/{\rm Fe}]\rangle=0.21\pm0.04$~dex. Isochrone fitting gives an age of $3.7\pm0.2$~Gyr, in good agreement with the chemical-clock estimate of $3.72\pm1.42$~Gyr.

\item The systemic radial velocity of Berkeley~31 is $57.89\pm0.08$~km~s$^{-1}$. The cluster has a relatively low orbital eccentricity of $e\simeq0.11\pm0.01$ and reaches a maximum height of $Z_{\rm max}=592\pm92$~pc above the Galactic plane. Its present Galactocentric radius is $R_{\rm GC}\simeq 13.85\pm0.20$~kpc, while the guiding radius is $R_{\rm g}\simeq 12.29\pm0.29$~kpc.

\item The photometric analysis of 189 MS stars gives an unresolved binary fraction of $0.38\pm0.04$ for systems with mass ratios of $q\gtrsim0.5$. Since the adopted method becomes less sensitive to lower mass-ratio systems, this value should be regarded as a lower limit to the intrinsic binary fraction. Spectroscopic monitoring will be useful for identifying the orbital properties of these systems and for constraining the binary population more completely.

\item We identified 62 BSS and two YSS candidates among the probable members of Berkeley~31. Of these, 48 BSSs and both YSSs are presented for the first time in this study. We constructed multi-wavelength SEDs for the full sample and investigated the 28 BSSs with available \textit{Swift}/UVOT measurements for evidence of additional hot components.

\item Two BSSs show ultraviolet excesses in at least two \textit{Swift}/UVOT bands. Two-component SED fits indicate hot compact companions with temperatures of approximately 45,000~K and 16,500~K. The hotter component is consistent with a hot WD, while the cooler component is consistent with a pre-ELM WD. Together with our previous study, the number of BSSs in Berkeley~31 with evidence for hot companions is now five, including one hot WD and four pre-ELM WD candidates.

\item The comparison between the present Galactocentric radius, guiding radius, and chemical birth radius suggests radial redistribution during the cluster's Galactic evolution. The chemical birth radius is $R_{\rm b}\simeq11.15\pm0.50$~kpc, smaller than the present guiding radius of $12.29\pm0.29$~kpc. This difference indicates an outward shift of the guiding center, while the larger difference between $R_{\rm b}$ and the present $R_{\rm GC}$ also reflects the cluster's current orbital phase. Thus, both radial migration and orbital motion contribute to the present radial position of Berkeley~31.

\end{itemize}

The results provide information on both the chemical and dynamical history of Berkeley~31. The difference between the present guiding radius and the estimated birth radius is consistent with a change in the cluster's guiding center, while the difference between the current Galactocentric radius and the birth radius also depends on the cluster's present orbital position. The hot compact companions identified in some BSSs are consistent with a past mass-transfer phase and suggest that binary evolution may have contributed to the formation of part of the BSS population. Multi-epoch, high-resolution spectroscopic observations are needed to confirm the binary nature of these systems and, where possible, determine their orbital properties. Additional UV observations would be useful for BSSs without suitable \textit{Swift}/UVOT data, since the lack of UV observations does not rule out the presence of a hot companion. A larger spectroscopic sample would also help improve the chemical characterization of Berkeley~31 and provide a better estimate of its birth radius. Further studies of OCs with different ages, metallicities, and Galactic positions could help place the results for Berkeley~31 in a broader context.

\begin{acknowledgments}
We thank the anonymous referee for their insightful and constructive suggestions that significantly improved the paper. Ing-Guey Jiang acknowledges support from the National Science and Technology Council (NSTC), Taiwan, under grants NSTC 113-2112-M-007-030 and NSTC 114-2112-M-007-029. This research, which forms part of the PhD thesis of Deniz Cennet Çınar, was supported by a doctoral scholarship from the Scientific and Technological Research Council of Türkiye (TÜBİTAK) through the BİDEB 2211 - National PhD Scholarship Programs. We also made use of NASA's Astrophysics Data System. This publication makes use of VOSA, developed under the Spanish Virtual Observatory (https://svo.cab.inta-csic.es) project funded by MCIN/AEI/10.13039/501100011033 through grant PID2023-146210NB-I00 and partially updated with funding from the European Union's Horizon 2020 Research and Innovation Programme under Grant Agreement nº 776403 (EXOPLANETS-A). The VizieR and SIMBAD databases at CDS, Strasbourg, France, were invaluable for this work, as were data from the European Space Agency (ESA) mission Gaia\footnote{https://www.cosmos.esa.int/gaia}, processed by the Gaia Data Processing and Analysis Consortium (DPAC)\footnote{https://www.cosmos.esa.int/web/gaia/dpac/consortium}. National institutions have funded DPAC, particularly those participating in the \emph{Gaia} Multilateral Agreement. Based on data obtained from the ESO Science Archive Facility with DOI: https://doi.org/10.18727/archive/25. We acknowledge the use of observations obtained with the Samuel Oschin 48-inch Telescope at Palomar Observatory as part of the Zwicky Transient Facility (ZTF) project. ZTF is supported by the National Science Foundation under Grant No. AST-1440341 and by a collaboration including Caltech, IPAC, the Weizmann Institute for Science, the Oskar Klein Center at Stockholm University, the University of Maryland, the University of Washington, Deutsches Elektronen-Synchrotron (DESY), Humboldt University, Los Alamos National Laboratory, the TANGO Consortium of Taiwan, the University of Wisconsin--Milwaukee, and Lawrence Berkeley National Laboratory.
\end{acknowledgments}
\facilities{Gaia, Swift, PS1, ZTF ,CTIO:2MASS}

\software{\texttt{galpy} \citep{Bovy2015}}

\appendix
\twocolumngrid

\section{Gaia DR3 Data Selection and Quality Criteria}
\label{app:data_quality}
We constructed the Gaia~DR3 catalog by querying sources within a $20$~arcmin radius of the adopted center of Berkeley~31, with an initial magnitude limit of $G<23$~mag. The query included the astrometric and photometric parameters required for the membership and photometric analyses. Only sources with five-parameter astrometric solutions, selected using the \texttt{astrometric\_params\_solved} parameter, were retained for membership determination, as reliable parallax and proper-motion measurements are needed for this analysis. The full query and selection criteria are provided for reproducibility.

We examined the photometric uncertainties as a function of $G$ magnitude and found that they increase toward fainter sources, particularly for the $G_{\rm BP}-G_{\rm RP}$ color. We therefore adopted $G \leq 20.5$~mag as the limiting magnitude for the CMD analysis to reduce the impact of large photometric uncertainties. The resulting catalog was used for the membership analysis described in Section~\ref{sec:membership} and the subsequent photometric analysis of Berkeley~31.

\section{Swift/UVOT Flux Measurements}\label{app:swift_data}

The \textit{Swift}/UVOT photometric measurements used in this study are listed in Table~\ref{tab:swift_uvot_fluxes}. We cross-matched the Gaia DR3 positions of the Berkeley~31 BSS candidates with the UVOT-OC catalog of \citet{Siegel2019}. A maximum separation of 1 arcsec was adopted for the positional match. The UVOT-OC catalog provides calibrated point-source photometry in the \textit{UVW2}, \textit{UVM2}, and \textit{UVW1} bands, obtained from the \textit{Swift}/UVOT observations of Galactic OCs. Source detection and photometry in the catalog were performed using DAOPHOT, followed by the standard UVOT calibration procedures, including the coincidence-loss and large-scale sensitivity corrections \citep{Siegel2019}. For the matched sources, we used the reported UVOT fluxes and their uncertainties in the SED analysis. The table gives the Gaia DR3 source IDs, equatorial coordinates, and the corresponding UVOT flux measurements. 

\begin{table*}[ht]
\centering
\scriptsize
\caption{\textit{Swift}/UVOT photometric measurements for the Berkeley~31 BSS candidates}. 
\label{tab:swift_uvot_fluxes}
\renewcommand{\arraystretch}{0.95}
\setlength{\tabcolsep}{4pt}
\begin{tabular}{c c c c c c c}
\hline
BSS ID & Gaia DR3 ID & $\alpha$ & $\delta$ &
\multicolumn{3}{c}{Observed Flux (erg s$^{-1}$ cm$^{-2}$ \AA$^{-1}$)} \\
\cline{5-7}
& & (degree) & (degree) & \textit{UVW}2 & \textit{UVM}2 & \textit{UVW}1 \\
\hline

01 & 3157142017325494656 & 104.310560 & 8.145330 &
-- &
7.983E-17 $\pm$ 8.602E-18 &
1.353E-16 $\pm$ 4.610E-18 \\

02 & 3157236579622368000 & 104.402613 & 8.223951 &
3.021E-16 $\pm$ 6.122E-18 &
3.563E-16 $\pm$ 6.562E-18 &
6.229E-16 $\pm$ 6.884E-18 \\

03 & 3157237885292374912 & 104.485478 & 8.227897 &
1.626E-16 $\pm$ 7.037E-18 &
2.612E-16 $\pm$ 9.141E-18 &
4.529E-16 $\pm$ 1.043E-17 \\

04 & 3157238808706887680 & 104.504155 & 8.299490 &
9.849E-17 $\pm$ 3.538E-18 &
1.426E-16 $\pm$ 4.072E-18 &
2.295E-16 $\pm$ 6.130E-18 \\

05 & 3157239976938373504 & 104.300830 & 8.214860 &
2.400E-17 $\pm$ 2.232E-18 &
2.362E-17 $\pm$ 2.285E-18 &
7.734E-17 $\pm$ 2.849E-18 \\

06 & 3157241080748060928 & 104.421720 & 8.256250 &
1.261E-16 $\pm$ 5.341E-18 &
1.400E-16 $\pm$ 5.029E-18 &
3.281E-16 $\pm$ 6.346E-18 \\

07 & 3157241454409817728 & 104.404817 & 8.283106 &
8.507E-17 $\pm$ 6.582E-18 &
7.823E-17 $\pm$ 5.836E-18 &
2.103E-16 $\pm$ 1.743E-17 \\

08 & 3157241493064902912 & 104.414398 & 8.291266 &
5.508E-17 $\pm$ 3.754E-18 &
5.447E-17 $\pm$ 3.612E-18 &
1.489E-16 $\pm$ 5.073E-18 \\

09 & 3157241523126691328 & 104.407218 & 8.289976 &
3.878E-17 $\pm$ 2.822E-18 &
2.851E-17 $\pm$ 2.626E-18 &
1.064E-16 $\pm$ 4.996E-18 \\

10 & 3157241523129294848 & 104.407380 & 8.294434 &
4.855E-17 $\pm$ 3.980E-18 &
-- &
1.788E-16 $\pm$ 1.416E-17 \\

11 & 3157241660565158528 & 104.442534 & 8.292120 &
9.768E-17 $\pm$ 4.138E-18 &
1.348E-16 $\pm$ 4.346E-18 &
2.062E-16 $\pm$ 5.699E-18 \\

12 & 3157242244681218944 & 104.374562 & 8.305973 &
3.550E-17 $\pm$ 2.256E-18 &
4.166E-17 $\pm$ 3.607E-18 &
1.008E-16 $\pm$ 7.058E-18 \\

13 & 3157242794437080192 & 104.414189 & 8.339619 &
-- &
2.835E-17 $\pm$ 2.507E-18 &
8.904E-17 $\pm$ 3.362E-18 \\

14 & 3157243004892739456 & 104.408358 & 8.350141 &
5.183E-17 $\pm$ 2.960E-18 &
7.062E-17 $\pm$ 3.252E-18 &
1.238E-16 $\pm$ 3.079E-18 \\

15 & 3157243206753998976 & 104.415060 & 8.375595 &
1.310E-16 $\pm$ 3.862E-18 &
1.531E-16 $\pm$ 4.371E-18 &
3.287E-16 $\pm$ 5.752E-18 \\

16 & 3157249155285881472 & 104.460890 & 8.392985 &
5.279E-16 $\pm$ 1.410E-17 &
4.850E-16 $\pm$ 1.206E-17 &
8.953E-16 $\pm$ 1.897E-17 \\

17 & 3157254923424787200 & 104.386276 & 8.379921 &
3.825E-16 $\pm$ 5.988E-18 &
4.090E-16 $\pm$ 8.665E-18 &
5.420E-16 $\pm$ 9.485E-18 \\

18 & 3157255133880388352 & 104.344806 & 8.385819 &
1.409E-16 $\pm$ 4.802E-18 &
1.619E-16 $\pm$ 4.176E-18 &
2.987E-16 $\pm$ 5.502E-18 \\

19 & 3157241115107807872 & 104.403860 & 8.257080 &
5.121E-16 $\pm$ 8.491E-18 &
5.477E-16 $\pm$ 7.063E-18 &
6.625E-16 $\pm$ 7.322E-18 \\

20 & 3157241282608500096 & 104.431740 & 8.281000 &
2.135E-16 $\pm$ 5.702E-18 &
2.465E-16 $\pm$ 5.448E-18 &
3.087E-16 $\pm$ 5.972E-18 \\

21 & 3157241355625967744 & 104.400720 & 8.274750 &
1.547E-15 $\pm$ 1.567E-17 &
1.618E-15 $\pm$ 1.490E-17 &
1.883E-15 $\pm$ 2.081E-17 \\

22 & 3157238263249488256 & 104.453034 & 8.267360 &
2.429E-16 $\pm$ 7.829E-18 &
3.237E-16 $\pm$ 5.963E-18 &
4.686E-16 $\pm$ 6.042E-18 \\

23 & 3157241351327999104 & 104.411767 & 8.274061 &
1.086E-16 $\pm$ 3.601E-18 &
1.358E-16 $\pm$ 3.753E-18 &
1.780E-16 $\pm$ 4.754E-18 \\

24 & 3157241561783738496 & 104.416514 & 8.303800 &
1.745E-16 $\pm$ 4.982E-18 &
1.841E-16 $\pm$ 4.408E-18 &
3.451E-16 $\pm$ 5.722E-18 \\

25 & 3157242283338251648 & 104.390768 & 8.303359 &
1.777E-16 $\pm$ 4.584E-18 &
2.084E-16 $\pm$ 3.647E-18 &
2.790E-16 $\pm$ 5.653E-18 \\

26 & 3157241523126689536 & 104.406420 & 8.288290 &
4.684E-16 $\pm$ 1.122E-17 &
4.909E-16 $\pm$ 9.494E-18 &
6.206E-16 $\pm$ 1.372E-17 \\

27 & 3157241557486428160 & 104.418610 & 8.295890 &
3.856E-16 $\pm$ 9.590E-18 &
4.053E-16 $\pm$ 6.719E-18 &
4.734E-16 $\pm$ 7.848E-18 \\

28 & 3157241527424009216 & 104.400688 & 8.290409 &
1.952E-16 $\pm$ 5.755E-18 &
2.178E-16 $\pm$ 5.217E-18 &
3.597E-16 $\pm$ 7.289E-18 \\

\hline
\end{tabular}
\end{table*}

\section{Single SED Fitting Results}
\label{app:single_sed}

\begin{figure}[H]
    \centering   \includegraphics[width=0.9\linewidth]{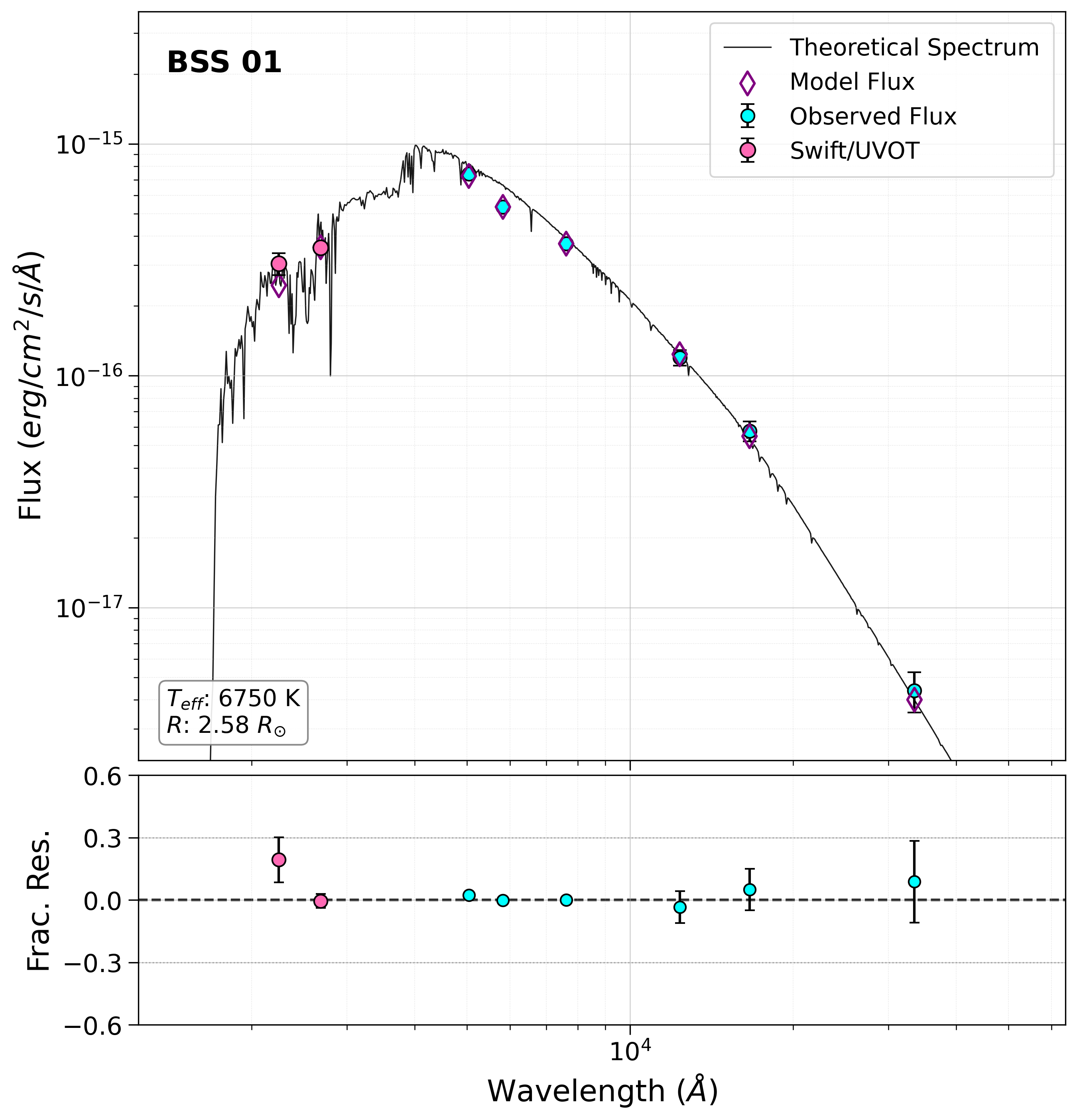}
    \caption{Single-component SED fit for BSS~01 in Berkeley~31. The upper panel shows the observed photometric fluxes (cyan circles) together with the best-fitting model spectrum (black curve). \textit{Swift}/UVOT measurements are marked by pink circles, while purple diamonds represent the corresponding synthetic fluxes. The lower panel shows the fractional residuals between the observed and fitted fluxes.}
    \label{fig:b31_sed}

\figsetstart
\figsetnum{1}
\figsettitle{The complete set of 16 SED figures is provided as online-only supplementary material.}

\figsetgrpstart
\figsetgrpnum{1.1}
\figsetgrptitle{SED for BSS 1}
\figsetplot{Figures/BSS_01.png}
\figsetgrpnote{SED for BSS 1}
\figsetgrpend

\figsetend
    
\end{figure}

The results of the single-component SED fitting for the Berkeley~31 BSS candidates are presented in Table~\ref{tab:single_sed_results}. The fitted parameters are the effective temperature ($T_{\rm eff}$), radius ($R$), and luminosity ($L$), together with the scaling factor, the number of photometric points used in the fit ($N_{\rm fit}$), the reduced chi-square ($\chi^2_\nu$), and the goodness-of-fit metrics vgf and vgf$_b$. The corresponding SEDs for the 16 BSSs with sufficient multi-wavelength photometric coverage are shown in Figure~\ref{fig:b31_sed}. The individual SED plots are provided electronically as supplementary material.

\begin{table*}
\centering
\scriptsize
\caption{Single-component SED fitting results for the Berkeley~31 BSS candidates.}
\label{tab:single_sed_results}
\renewcommand{\arraystretch}{0.95}
\setlength{\tabcolsep}{4pt}
\begin{tabular}{cccccccccccc}
\hline
BSS ID & $\alpha$ (hh:mm:ss) & $\delta$ (dd:mm:ss) &
$T_{\rm eff}$ (K) & $R$ ($R_\odot$) & $L$ ($L_\odot$) &
Scaling Factor & $N_{\rm fit}$ & $\chi^2_\nu$ & vgf & vgf$_b$ \\
\hline

BSS 01 & 06:57:14.53 & +08:08:43.2 &
6750$\pm$123 & 2.58$\pm$0.10 & 12.51$\pm$0.05 &
4.38E-23 & 8 & 1.67 & 1.04 & 0.77 \\

BSS 02 & 06:57:36.63 & +08:13:26.2 &
7000$\pm$135 & 4.13$\pm$0.15 & 37.56$\pm$0.01 &
1.67E-22 & 14 & 17.84 & 3.42 & 0.44 \\

BSS 03 & 06:57:56.51 & +08:13:40.4 &
6750$\pm$175 & 2.72$\pm$0.10 & 13.78$\pm$0.03 &
1.51E-22 & 16 & 33.72 & 6.87 & 0.84 \\

BSS 04 & 06:58:01.00 & +08:17:58.2 &
6750$\pm$155 & 2.99$\pm$0.11 & 16.84$\pm$0.02 &
8.09E-23 & 16 & 37.15 & 10.88 & 0.92 \\

BSS 05 & 06:57:12.20 & +08:12:53.5 &
6000$\pm$125 & 2.71$\pm$0.11 & 8.60$\pm$0.02 &
6.54E-23 & 10 & 2.80 & 2.65 & 0.72 \\

BSS 06 & 06:57:41.21 & +08:15:22.5 &
6500$\pm$168 & 4.38$\pm$0.17 & 31.08$\pm$0.01 &
1.39E-22 & 16 & 20.17 & 4.88 & 0.61 \\

BSS 08 & 06:57:39.46 & +08:17:28.6 &
6000$\pm$123 & 1.81$\pm$0.08 & 3.85$\pm$0.00 &
6.81E-23 & 16 & 20.65 & 8.77 & 3.62 \\

BSS 09 & 06:57:37.73 & +08:17:23.9 &
6500$\pm$120 & 2.09$\pm$0.08 & 7.22$\pm$0.01 &
6.17E-23 & 14 & 43.87 & 10.10 & 4.49 \\

BSS 10 & 06:57:37.77 & +08:17:40.0 &
6750$\pm$125 & 1.04$\pm$0.04 & 2.02$\pm$0.00 &
3.72E-23 & 15 & 22.62 & 4.94 & 0.49 \\

BSS 11 & 06:57:46.21 & +08:17:31.6 &
6250$\pm$187 & 2.36$\pm$0.09 & 7.61$\pm$0.01 &
6.95E-23 & 15 & 23.29 & 2.83 & 1.20 \\

BSS 13 & 06:57:39.41 & +08:20:22.6 &
6750$\pm$157 & 4.09$\pm$0.15 & 31.44$\pm$0.02 &
1.59E-22 & 16 & 11.71 & 3.32 & 0.62 \\

BSS 14 & 06:57:50.61 & +08:23:34.7 &
7250$\pm$156 & 3.12$\pm$0.11 & 24.15$\pm$0.03 &
2.43E-22 & 16 & 50.05 & 5.81 & 0.47 \\

BSS 15 & 06:57:32.71 & +08:22:47.7 &
8000$\pm$125 & 2.24$\pm$0.07 & 18.71$\pm$0.01 &
6.49E-23 & 16 & 19.19 & 5.74 & 0.39 \\

BSS 16 & 06:57:22.75 & +08:23:08.9 &
7000$\pm$125 & 1.76$\pm$0.06 & 6.68$\pm$0.00 &
8.68E-23 & 14 & 29.97 & 2.24 & 0.30 \\

BSS 17 & 06:57:36.93 & +08:15:25.5 &
7500$\pm$156 & 1.54$\pm$0.10 & 6.70$\pm$0.70 &
7.84E-23 & 18 & 201.37 & 7.15 & 2.31 \\

BSS 18 & 06:57:12.20 & +08:12:53.5 &
10000$\pm$274 & 2.20$\pm$0.11 & 44.04$\pm$8.78 &
2.16E-22 & 25 & 11.19 & 7.15 & 1.48 \\

BSS 19* &  06:57:36.93         & +08:15:25.49        & 8000$\pm$167      & 1.25$\pm$0.06   & 5.74$\pm$1.10    & 3.39E-23       & 13            & 169.71 & 29.05 & 0.56    \\
BSS 20* & 06:57:43.62         & +08:16:51.60        & 10000$\pm$301     & 1.14$\pm$0.06   & 11.80$\pm$1.89   & 2.93E-23       & 20            & 4.94   & 4.88  & 0.93    \\
BSS 21* & 06:57:36.17         & +08:16:29.10        & 8000$\pm$174      & 1.50$\pm$0.07   & 8.35$\pm$1.25    & 9.08E-23       & 15            & 36.04  & 14.16 & 0.62    \\
BSS 22* &  06:57:48.73         & +08:16:02.50        & 7500$\pm$125      & 1.54$\pm$0.08   & 6.70$\pm$0.70    & 7.84E-23       & 18            & 22.48  & 7.99  & 2.31    \\
BSS 23* &06:57:38.82         & +08:16:26.62        & 7500$\pm$212      & 0.93$\pm$0.05   & 2.48$\pm$0.47    & 2.40E-23       & 14            & 2.53   & 2.52  & 0.45    \\
BSS 24* &  06:57:39.96         & +08:18:13.68        & 8000$\pm$125      & 2.57$\pm$0.13   & 23.92$\pm$2.40   & 7.77E-23       & 24            & 44.96  & 7.15  & 2.26    \\
BSS 25* & 06:57:33.78         & +08:18:12.09        & 7500$\pm$159      & 1.35$\pm$0.06   & 5.18$\pm$0.81    & 4.56E-23       & 15            & 76.70   & 12.38 & 0.51    \\

\hline
\end{tabular}
\begin{flushleft}
\scriptsize
$^{*}$ Results adopted from \citet{Cinar2026b}.
\end{flushleft}

\end{table*}


\bibliography{references}{}
\bibliographystyle{aasjournalv7}

\end{document}